\documentclass[11pt]{article}

\usepackage[letterpaper, left=1in, top=1in, right=1in, bottom=1in,nohead,includefoot]{geometry}
\usepackage{color,charter,graphicx,natbib,here,setspace,multirow,amsmath,amssymb,amsfonts,enumitem}
\usepackage[dvipsnames]{xcolor}
\usepackage{graphicx}
\usepackage{rotating}
\usepackage{multirow}
\usepackage{booktabs}
\usepackage{color}
\usepackage{setspace}
\usepackage{algorithm2e}
\usepackage{enumitem}
\usepackage[bottom]{footmisc}
\usepackage[colorlinks=true,urlcolor=blue,linkcolor=blue,citecolor=blue]{hyperref}
\usepackage{bigstrut}
\usepackage{threeparttable}
\usepackage{dcolumn}
\usepackage{amsmath}
\usepackage{longtable}
\usepackage{amssymb}
\usepackage{lscape }

\pdfoutput=1
\usepackage{eqparbox}

\usepackage[title,titletoc]{appendix}
\makeatletter

\renewenvironment{appendices}{%
    \begin{oldappendices}%
    \renewcommand{\thefigure}{\ifnum \c@section>\z@ \thesection.\fi\@arabic\c@figure}%
    \@addtoreset{figure}{section}%
    \renewcommand{\thetable}{\ifnum \c@section>\z@ \thesection.\fi\@arabic\c@table}%
    \@addtoreset{table}{section}%
}{%
    \end{oldappendices}%
}\makeatother

\usepackage{natbib}
\let\natbibcitet\citet
\renewcommand\citet{\bibpunct{(}{)}{,}{a}{,}{,}\natbibcitet}
\let\natbibcitep\citep
\renewcommand\citep{\bibpunct{(}{)}{;}{a}{,}{;}\natbibcitep}
\newcommand{\bi}{\begin{itemize}}
\newcommand{\ei}{\end{itemize}}
\newcommand{\be}{\begin{equation}}
\newcommand{\ee}{\end{equation}}
\defcitealias{ieo14}{IEO, 2014}

\long\def\symbolfootnote[#1]#2{\begingroup%
\def\thefootnote{\fnsymbol{footnote}}\footnote[#1]{#2}\endgroup}

\usepackage{sectsty}
\allsectionsfont{\normalsize}

\makeatletter

\@addtoreset{equation}{section}
\makeatother

\usepackage{caption}
\usepackage[labelformat=simple]{subcaption}
\makeatletter\let\p@subfigure\thefigure\makeatother

\usepackage{cleveref}
\crefname{chapter}{Chapter}{Chapters}
\crefname{section}{Section}{Sections}
\crefname{subsection}{Section}{Sections}
\crefname{subsubsection}{Section}{Sections}
\crefname{figure}{Figure}{Figures}
\crefname{table}{Table}{Tables}
\crefname{equation}{Equation}{Equations}
\crefname{appendix}{Appendix}{Appendices}

\newcolumntype{d}[1]{D{.}{.}{#1}}

\usepackage{authblk}
\usepackage{bm}

  \title{\bf \large Predicting crypto-currencies  using sparse non-Gaussian state space models}
  \author[1]{Christian Hotz-Behofsits}
  \author[2]{Florian Huber\thanks{Corresponding author: Florian Huber, Vienna University of Economics and Business. E-mail: \href{mailto:fhuber@wu.ac.at}{fhuber@wu.ac.at}. This paper, with minor editorial changes, is forthcoming in the Journal of Forecasting. The authors thank  Gregor Kastner, O. Skar, Belinda Haid, Jouchi Nakajima, and the participants of the 2017 NBP Workshop on Forecasting for helpful comments and suggestions. Financial support from the Czech Science Foundation, Grant 17-14263S, is gratefully acknowledged.}}
    \author[2]{Thomas O. Z\"{o}rner}
 \affil[1]{Vienna University of Economics and Business, Department of Marketing}
\affil[2]{Vienna University of Economics and Business, Department of Economics}

 \date{}


\def\equationautorefname~#1\null{%
  Eq.~(#1)\null
}
\def\equationautorefname~#1\null{
Eq.~(#1)\null
}

\begin{document}
\graphicspath{{Figures/}}
\maketitle

\begin{abstract}
\noindent
In this paper we forecast daily returns of crypto-currencies using a wide variety of different
econometric models. To capture salient features commonly observed in financial time series
like rapid changes in the conditional variance, non-normality of the measurement errors and sharply increasing trends, we develop a time-varying parameter VAR with t-distributed measurement errors and stochastic volatility. To control for overparameterization, we rely on the Bayesian literature on shrinkage priors that enables us to shrink coefficients associated with irrelevant predictors and/or perform model specification in a flexible manner. Using around one year of daily data we perform a real-time forecasting exercise and investigate whether any of the proposed models is able to  outperform the naive random walk benchmark. To assess the economic relevance of the forecasting gains  produced by the proposed models we moreover run a simple trading exercise.

\end{abstract}
\bigskip
\begin{tabular}{p{0.2\hsize}p{0.65\hsize}} 
\textbf{Keywords:}  & Stochastic volatility, t-distributed errors, Bitcoin, density forecasting\\
\end{tabular}\\
\smallskip
\begin{tabular}{p{0.2\hsize}p{0.4\hsize}}
\textbf{JEL Codes:} & C11, C32, E51, G12 \\
\end{tabular}
\vspace{0.6cm}

\bigskip

\newpage

\section{Introduction}
In the present paper we develop a non-Gaussian state space model to predict the price of three crypto-currencies. Taking a Bayesian stance enables us to introduce shrinkage into the modeling framework, effectively controlling for model and specification uncertainty within the general class of state space models. To control for potential outliers we  propose a time-varying parameter VAR model \citep{CogleySargent2005, primiceri2005time} with heavy tailed innovations\footnote{For a recent exposition on how to introduce flexible error distributions in VAR models with stochastic volatility, see \cite{mumtaz2017forecasting}.} as well as a stochastic volatility specification of the error variances. Since the literature on robust determinants of price movements in crypto-currencies is relatively sparse \citep[for an example, see,][]{cheah2015speculative}, we apply Bayesian shrinkage priors to decide whether using information from a set of potential predictors improves predictive accuracy.

The recent price dynamics of various crypto-currencies point towards a set of empirical key features an appropriate modeling strategy should accommodate. First, conditional heteroscedasticity appears to be an important regularity commonly observed \citep{chu2017garch}. This implies that volatility is changing over time with persistent manner. If this feature is neglected, predictive densities are either too wide (during tranquil times) or too narrow (in the presence of tail events, i.e. pronounced movements in the price of a given asset).\footnote{Controlling for heteroscedasticity in macroeconomic and financial data proves to be an important task when it comes to prediction, see \cite{clark2011real, clark2015macroeconomic, huber2017adaptive}.} Second, the conditional mean of the process is changing. This implies that, within a standard regression framework,  the relationship between an asset price and a set of exogenous covariates is time-varying. In the case of various crypto-currencies this could be due to changes in the degree of adoption of institutional and/or private investors, regulatory changes, issuance of additional crypto-currencies or general technological shifts \citep{bohme2015bitcoin}. Thus, it might be necessary to allow for such shifts by means of time-varying regression coefficients. Third, and finally, a rather strong degree of co-movement between various crypto-currencies \citep[see][]{urquhart2017price}. In our paper, we consider Bitcoin, Ethereum and Litecoin, three popular choices. All three of them tend to be strongly correlated with each other, implying that a successful econometric framework should incorporate this information.

The goal of this paper is to systematically assess how different empirically relevant forecasting models perform when used to predict daily changes in the price of Bitcoin, Ethereum and Litecoin. The models considered include a wide range of univariate and multivariate models that are flexible along several dimensions. We consider  vector autoregressions that feature drifting parameters as well as time-varying error variances. To cope with the curse of dimensionality we introduce recent shrinkage priors \citep[see][]{feldkircher2017sophisticated} and a flexible specification for the law of motion of the regression parameters \citep{huber2017new}. In addition, we introduce a heavy tailed measurement error distribution to capture potential outlying observations \citep[see, among others,][]{carlin1992monte, geweke2001bayesian}.

We jointly forecast the three crypto-currencies considered by using daily data from October 2016 to October 2017, with the last 160 days being used as a hold-out period. In a forecasting comparison, we find that time-varying parameter VARs with some form of shrinkage perform well, beating univariate benchmarks like the AR(1) model with stochastic volatility (SV) as well as a random walk with SV. Constant parameter VARs tend to be inferior to their counterparts that feature time-varying parameters, but still prove to be relevant competitors. Especially during days which are characterized by large price changes, controlling for heteroscedasticity in combination with a flexible error variance covariance structure pays off in terms of predictive accuracy. These findings are generally corroborated by considering probability integral transforms, showing that more flexible models lead to better calibrated predictive distributions. Moreover, a trading exercise provides a comparable picture. Models that perform well in terms of predictive likelihoods also tend to do well when used to generate trading signals.

The remainder of this paper is structured as follows. Section \ref{sec:keyfeatures} provides an overview of the data as well as empirical key features of the three crypto-currencies considered. Moreover, this section details how the additional explanatory variables are constructed. Section \ref{sec:model} introduces the econometric framework adopted, providing a brief discussion of the model as well as the Bayesian prior setup and posterior simulation. Section \ref{sec:forecasting} presents the empirical forecasting exercise while Section \ref{sec:trading} focuses on applying the proposed models to perform portfolio allocation tasks. Finally, the last section summarizes and concludes the paper.

\section{Empirical key features}\label{sec:keyfeatures}
In this section we first identify important empirical key features of crypto-currencies and then propose a set of covariates that aim to explain the low to medium frequency behavior of the underlying price changes.

For the present paper, we  focus on the daily change in the log price of Bitcoin, Ethereum and Litecoin. To explain movements in the price of the three crypto-currencies considered, we include information on equity prices (measured through the log returns of the S\&P500 index), the relative number of search queries for each respective crypto-currency from Google trends, the number of English Wikipedia page views as well as the difference between the weekly cumulative price trend from common mining hardware and similar, but mining-unsuitable,  GPU-related products to capture the effect of supply-side factors.

The data spans the period from 26th November 2016 to 3rd October 2017, yielding a panel of 316 daily oberservations. Bitcoin, Ethereum and Litecoin closing prices are taken from a popular crypto-currency meta-platform.\footnote{For more information, see \url{coinmarketcap.com}.} They originate from major crypto exchanges and are averaged according to their daily trading volume. Furthermore, alternative financial investments are represented by the S\&P500 indices daily closing prices. Additionally, demand-side predictors like the relative number of world-wide search operations from Google trends and the number of  Wikipedia page views (in english) are used. Because large-scale crypto-currency mining impacts supply and prices for the required equipment at the same time, hardware price trends are utilized to express changes in supply. To capture these effects, we gather GPU prices from Amazon's bestseller lists and extract the price trend of common mining hardware. We construct this predictor by computing the difference between the weekly cumulative price trend from common mining hardware (e.g., AMD Radeon RX 480 graphic cards) and similar, but GPU-related products that are unsuitable for mining activities (e.g., a AMD Radeon R5 230 graphics card).

\begin{figure}[t]
\begin{minipage}[]{1\linewidth}
\centering
\textbf{Bitcoin}
\end{minipage}
\begin{minipage}[]{.45\linewidth}
\subcaption*{Log returns}
\centering
\includegraphics[scale=.2, trim=20 35 30 20, clip]{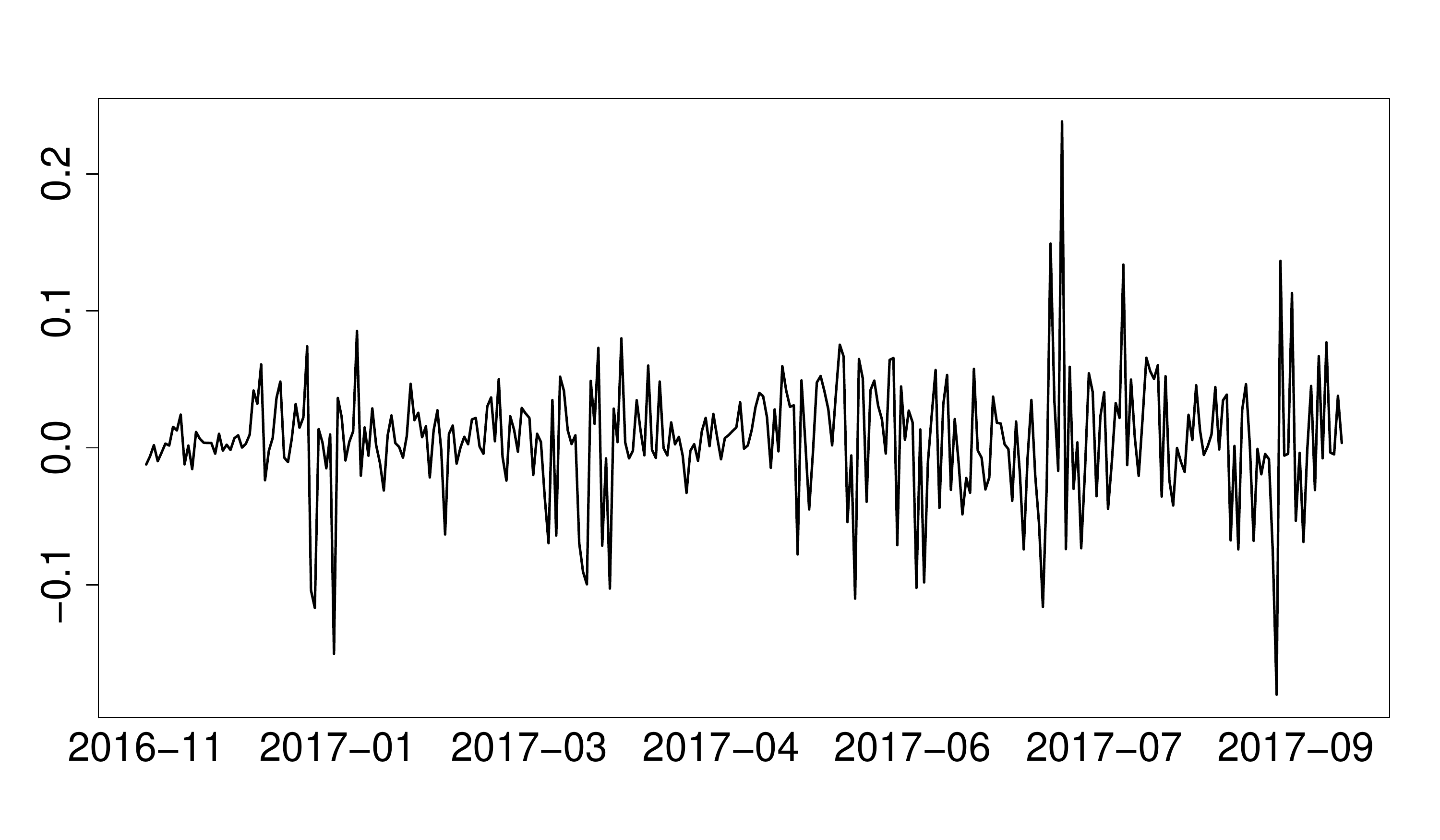}
\end{minipage}
\begin{minipage}[]{.45\linewidth}
\subcaption*{Squared log returns}
\centering
\includegraphics[scale=.2, trim=20 35 30 20, clip]{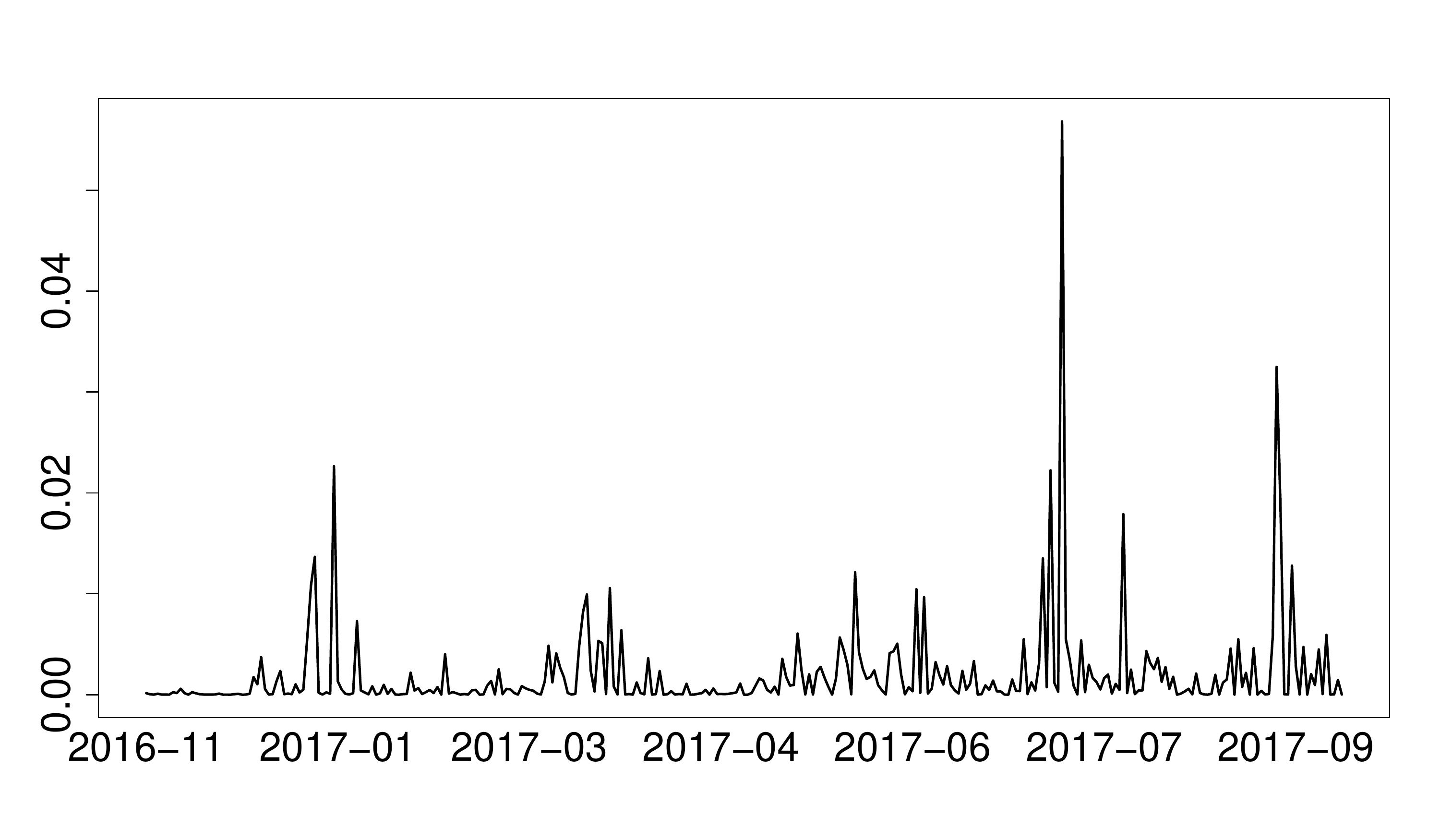}
\end{minipage}\\
\begin{minipage}[]{1\linewidth}
\centering
\textbf{Litecoin}
\end{minipage}
\begin{minipage}[]{.45\linewidth}
\centering
\includegraphics[scale=.2, trim=20 35 30 20, clip]{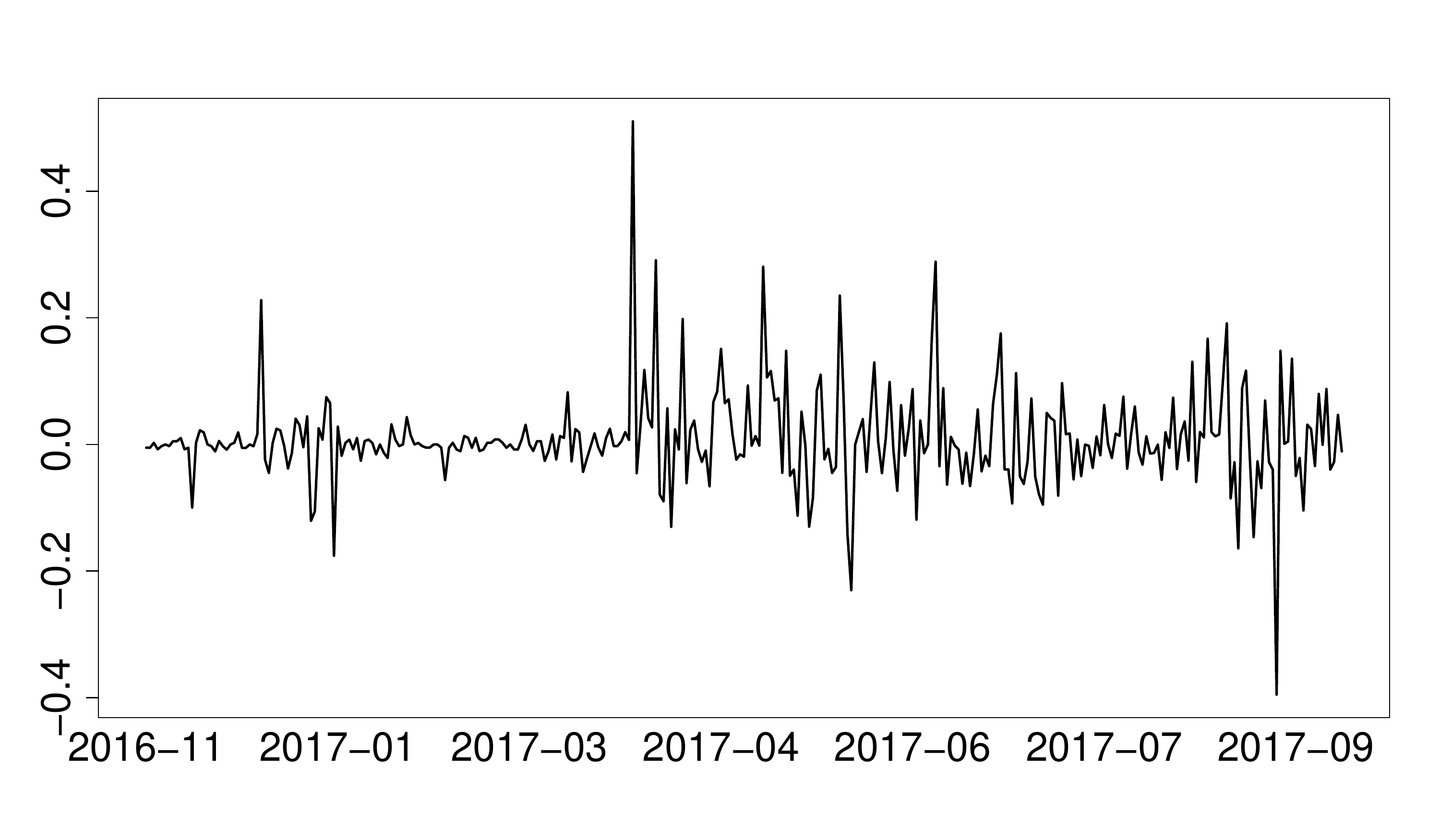}
\end{minipage}
\begin{minipage}[]{.45\linewidth}
\centering
\includegraphics[scale=.2, trim=20 35 30 20, clip]{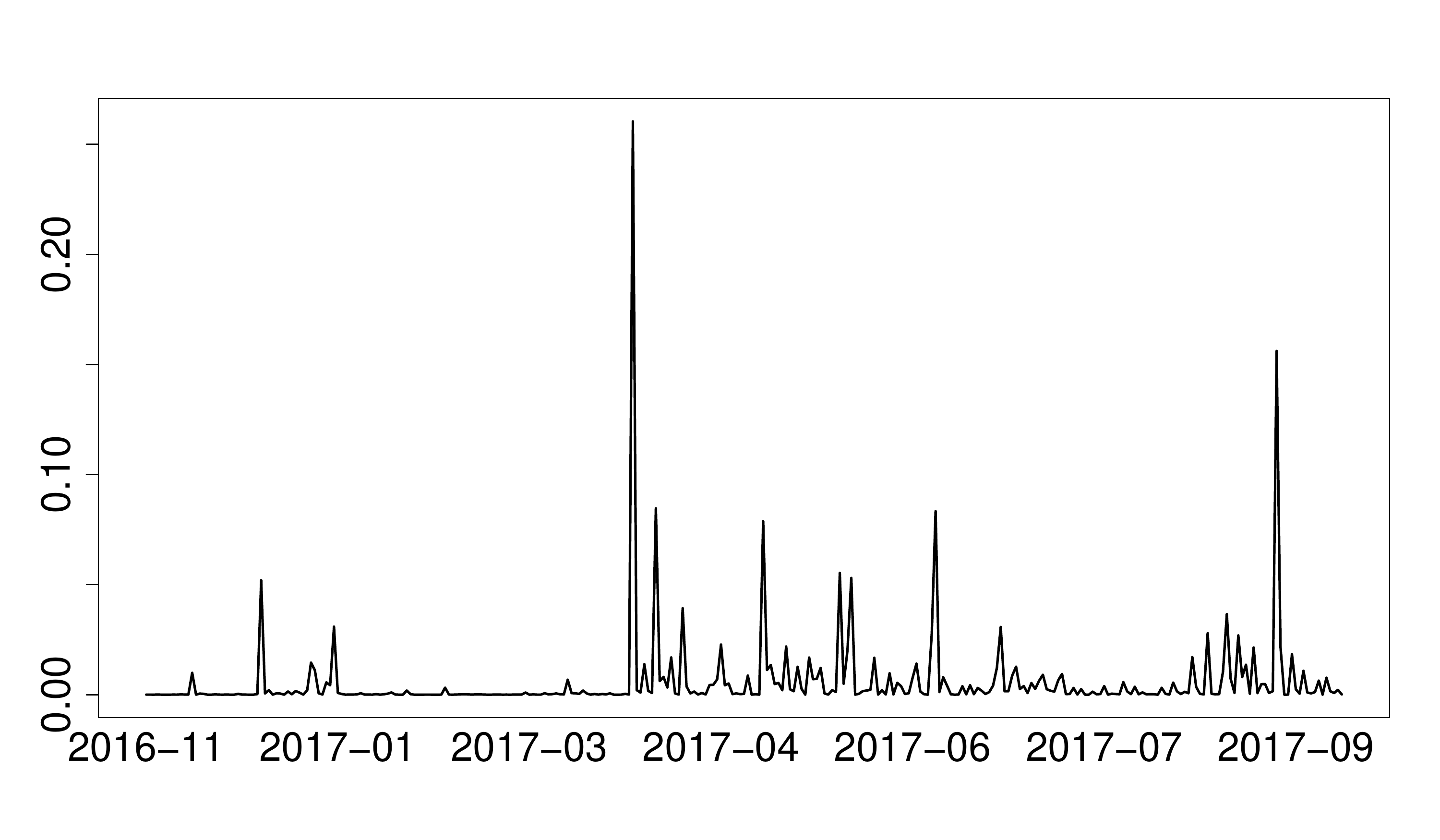}
\end{minipage}\\
\begin{minipage}[]{1\linewidth}
\centering
\textbf{Ethereum}
\end{minipage}
\begin{minipage}[]{.45\linewidth}
\centering
\includegraphics[scale=.2, trim=20 35 30 20, clip]{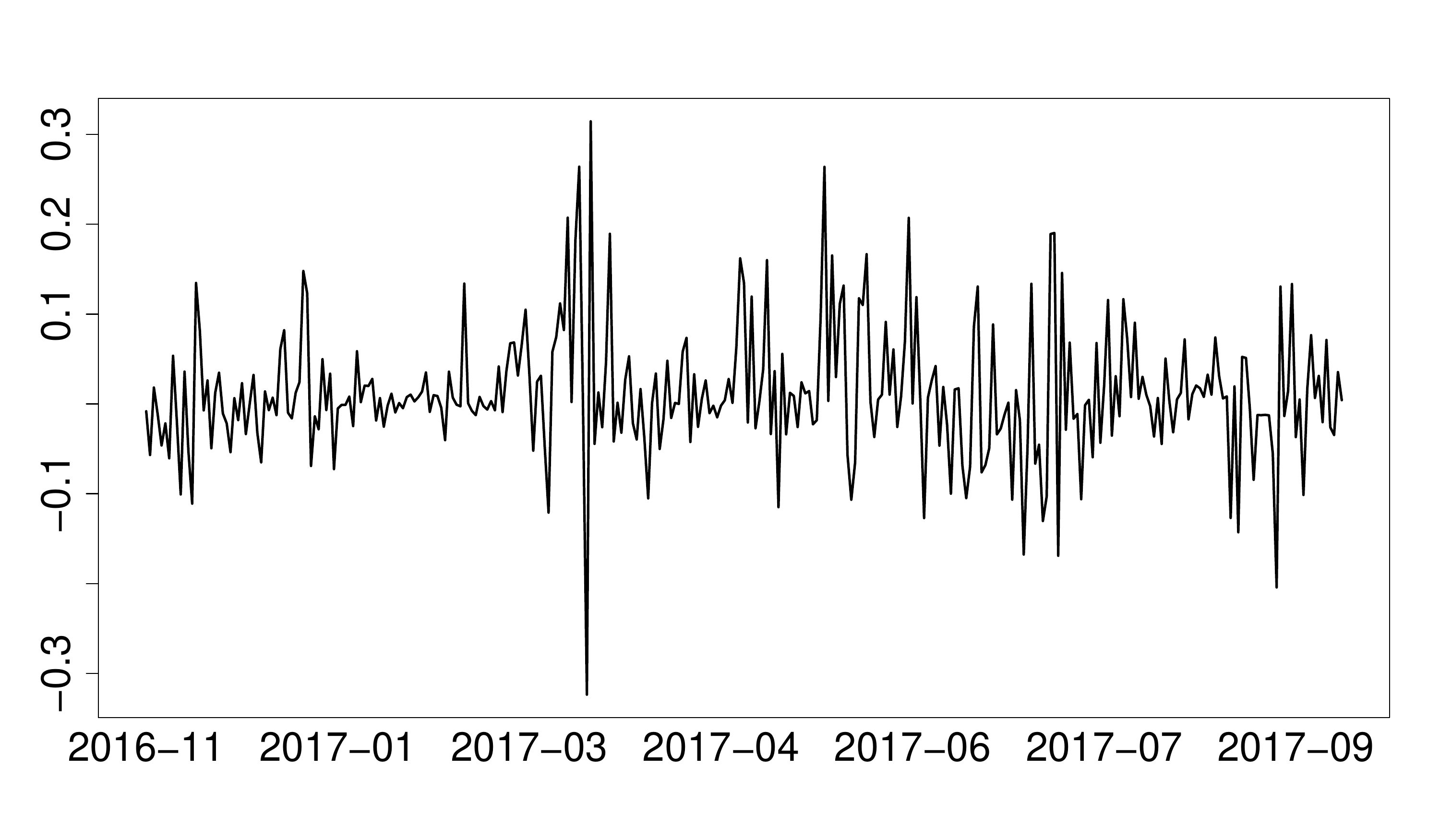}
\end{minipage}
\begin{minipage}[]{.45\linewidth}
\centering
\includegraphics[scale=.2, trim=20 35 30 20, clip]{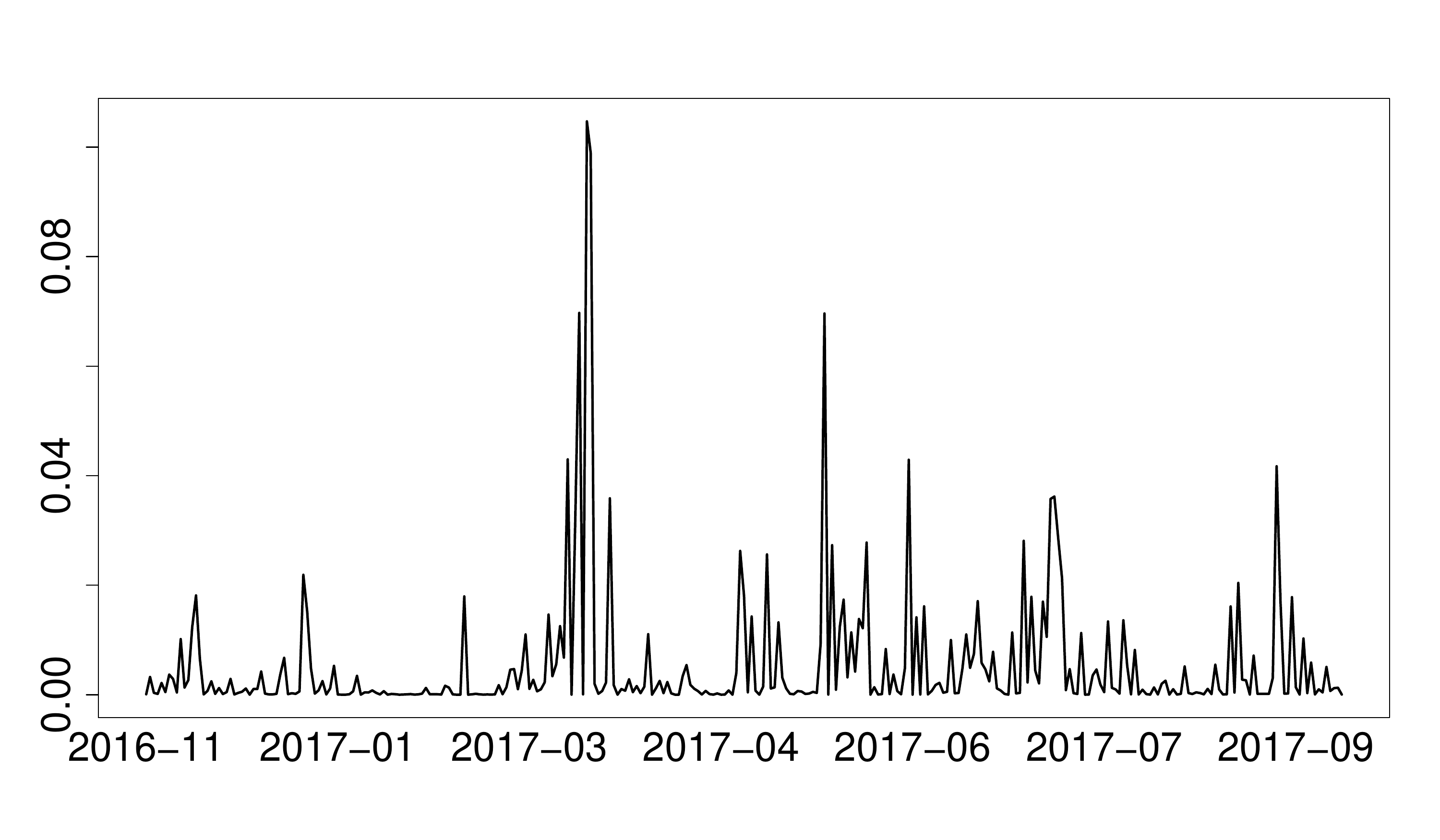}
\end{minipage}
\caption{Data overview: logarithmic returns and squared logarithmic returns} \label{fig:data}
\end{figure}

To provide additional information on the recent behavior of crypto-currencies, \autoref{fig:data} presents the log returns (left panel) as well as the squared log returns (right panel) for all three currencies under scrutiny.
At least two features are worth emphasizing. First, notice that in the first part of the sample (i.e. the end of 2016 and the beginning of 2017), price changes have been comparatively small. This can be seen in both panels of the figure and for Bitcoins and Litecoins. For Ethereum, the pattern is slightly different but we still observe a general increase in variation during the second part of 2017.

Second, the degree of co-movement between the three currencies increased markedly in 2017, where most major peaks and troughs coincide. This carries over to the squared returns, where we find that especially the sharp increase in volatility in September 2017 was common to all three currencies considered.

These two empirical regularities suggest that the proposed model should be able to capture co-movement between Bitcoin, Litecoin and Ethereum prices as well as changes in the first moment of the sampling density. Moreover, the right panel indicates that large shocks appear to be quite common, calling for a flexible error distribution that allows for heteroscedasticity.

\begin{figure}[h]
\centering
    \includegraphics[width=.75\textwidth]{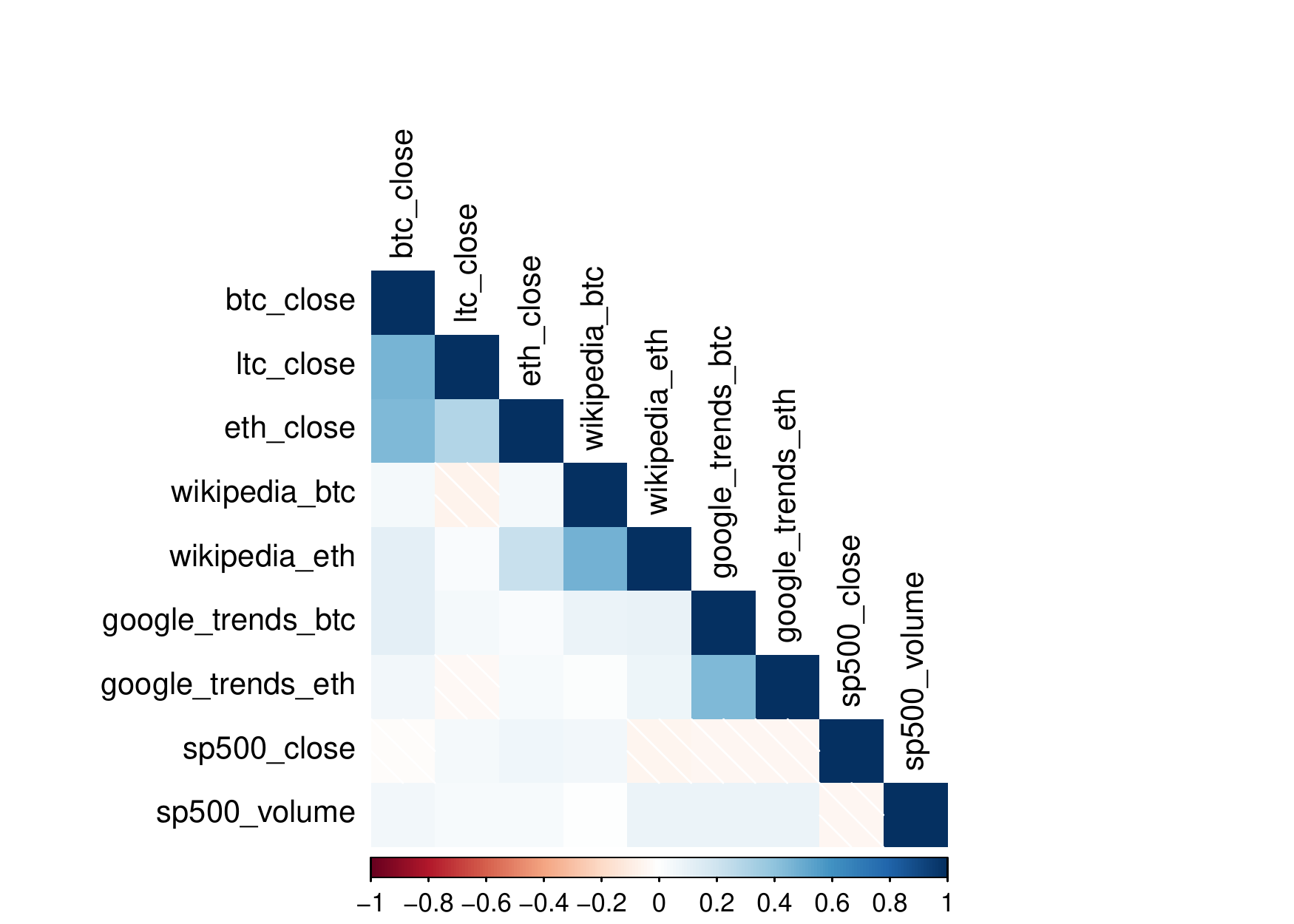}
    \caption{Lower Cholesky factor of the empirical correlation matrix of the dataset used.}\label{fig:corrmat}
\end{figure}

In order to provide further information on the amount of co-movement in our dataset, \autoref{fig:corrmat} shows a heatmap of the lower Cholesky factor of the empirical correlation matrix of the nine time series included.

The upper part of the figure reveals that all three assets display a pronounced degree of co-movement. This indicates that each individual time series might carry important information on the behavior of the remaining two time series, pointing towards the necessity to control for this empirical regularity. For the remaining factors we do find non-zero correlation but these correlations appear to be rather muted. Nevertheless, we conjecture that the set of fundamentals above should be a reasonable starting point to explain movements in the price of crypto-currencies.

\section{Econometric framework}\label{sec:model}
\subsection{A multivariate state space model}
To capture the empirical features of the three crypto-currencies, a flexible econometric model is needed. We assume that the three crypto-currencies as well as the additional covariates are stored in an $m$-dimensional vector $\{\bm{y}_t\}_{t=1}^T$ that follows a VAR($p$) model with time-varying coefficients,
\begin{equation}
\bm{y}_t = \bm{A}_{1t} \bm{y}_{t-1}+\dots +\bm{A}_{pt} \bm{y}_{t-p}+\bm{\varepsilon}_t, \label{eq: VAR}
\end{equation}
with $\bm{A}_{jt}$ (for $j=1,\dots, p$) being a set of $m \times m$-dimensional coefficient matrices and $\bm{\varepsilon}_t$ is a multivariate  vector of reduced form shocks  with  a time-varying variance covariance matrix,
\begin{equation}
\bm{\Sigma}_t = \bm{U}_t \bm{H}_t \bm{U}'_t.
\end{equation}
Hereby we let $\bm{U}_t$ be a lower uni-triangular matrix with $\text{diag}=\bm{\iota}_m$ and $\bm{\iota}_m$ being a $m$-dimensional vector of ones. Moreover, $\bm{H}_t$ is a diagonal matrix with typical diagonal element $[\bm{H}_t]_{jj}=e^{h_{jt}}$. The logarithmic volatilities are assumed to follow an AR(1) process,
\begin{equation}
h_{jt} = \mu_j + \rho_j (h_{jt-1}-\mu_j) + \varsigma_j \nu_{jt},\quad \nu_{jt} \sim \mathcal{N}(0,1). \label{eq:logvola}
\end{equation}
$\mu_j$ denotes the unconditional mean of the log-volatility process while $\rho_j$ and $\varsigma_j$ are the persistence and variance parameters, respectively.

Following \cite{carriero2015large} and \cite{feldkircher2017sophisticated} we rewrite \autoref{eq: VAR} as follows,
\begin{equation}
\tilde{\bm{U}}_t\bm{\varepsilon}_t =  \bm{\eta}_t,
\end{equation}
where $\tilde{\bm{U}}_t := \bm{U}^{-1}_t$  and $ \bm{\eta}_t$ is a vector of orthogonal shocks with a time-varying variance-covariance matrix.

Note that the $i$th equation (for $i>1$) of this system can be written as,
\begin{equation}
y_{it} = \bm{A}_{i\bullet} \bm{x}_t - \sum_{j=1}^{i-1}\tilde{u}_{ij} \varepsilon_{jt}+ \eta_{it}. \label{eq: eqbyeq}
\end{equation}
We let $\bm{x}_t=(\bm{y}'_{t-1}, \dots, \bm{y}'_{t-p})'$ be the stacked vector of covariates and   $\bm{A}_t=[\bm{A}_{1t}, \dots, \bm{A}_{pt}]$ is the $m \times mp$ matrix of stacked coefficients with $\bm{A}_{i\bullet, t}$ selecting the $i$th row of the matrix concerned.  \autoref{eq: eqbyeq} is a simple regression model with heteroscedastic innovations and the (negative) of the reduced form shocks of the preceding $i-1$ equations as additional regressors. In the case of $i=1$, \autoref{eq: eqbyeq} reduces to a simple univariate regression  with $\bm{x}_t$ as covariates. It proves to be convenient to rewrite \autoref{eq: eqbyeq} as follows
\begin{equation}
y_{it} = \bm{\beta}'_{it} \bm{z}_{it} +\eta_{it}, \label{eq: regressionsimple}
\end{equation}
where $\bm{\beta}_{it}=(\bm{A}_{i\bullet}, \tilde{u}_{i1}, \dots, \tilde{u}_{i i-1})'$ is a $k_i = m p + (i-1)$-dimensional vector of regression coefficients and  $\bm{z}_{it}=[\bm{x}'_t, -\varepsilon_{1t}, \dots, -\varepsilon_{i-1, t}]'$. One important implication of \autoref{eq: regressionsimple} is that the covariance parameters are effectively estimated in one step alongside the VAR coefficients.

We assume that $\bm{\beta}_{it}$ evolves according to a random walk process,
\begin{equation}
\bm{\beta}_{it} = \bm{\beta}_{it-1}+ \bm{e}_{it}. \label{eq: states}
\end{equation}
The shocks to the states $\bm{e}_{it}\sim \mathcal{N}(\bm{0}, \bm{\Theta}_i)$ follow a Gaussian distribution with diagonal variance-covariance matrix $\bm{\Theta}_i=\text{diag}(\vartheta_{i1}, \dots, \vartheta_{i k_i})$. To facilitate variable selection/shrinkage we follow \cite{fruhwirth2010stochastic, belmonte2014hierarchical, bitto2016achieving}  and rewrite the model given by Eqs. (\ref{eq: regressionsimple}) - (\ref{eq: states}) as follows,
\begin{align}
y_{it} &= \bm{\beta}'_{i0} \bm{z}_{it}+\tilde{\bm{\beta}}'_{it} \sqrt{\bm{\Theta}_i} \bm{z}_{it} +\eta_{it},\label{eq: obsNCP} \\
\tilde{\bm{\beta}}_{it} &= \tilde{\bm{\beta}}_{it-1}+\bm{\xi}_{it},\quad \bm{\xi}_{it} \sim \mathcal{N}(\bm{0}, \bm{I}_{k_i}),\\
\tilde{\bm{\beta}}_{i0}&=\bm{0}
\end{align}
The matrix $\sqrt{\bm{\Theta}_i}$ is a matrix square root such that $\bm{\Theta}_i=\sqrt{\bm{\Theta}_i}\sqrt{\bm{\Theta}_i}$ with typical element $\sqrt{\vartheta_{ij}}$ and $\tilde{{\beta}}'_{ij, t}$ the $j$th element of $\tilde{\bm{\beta}}'_{it}$ reads $(\beta_{ij,t}-\beta_{ij,0})/\sqrt{\vartheta_{ij}}$. This parameterization, labeled the non-centered parameterization, implies that the state innovation variances are moved into the observation equation (see \autoref{eq: obsNCP}) and treated as standard regression coefficients. Thus, if  $\sqrt{\vartheta_{i j}}=0$, the coefficient associated with the $j$th element in $\bm{z}_{it}$ is constant over time.

Up to this point we have remained silent on the distributional assumptions on the measurement errors. In what follows we depart from the literature on TVP-VARs and assume that the measurement errors are heavy tailed and follow a t-distribution.  This choice is based on  evidence in the  literature \citep{geweke1994bayesian, gallant1997estimation, jacquier2004bayesian} which calls for heavy tailed distributions when used to model daily financial market data. As can be seen in \autoref{fig:data}, we also observe multiple outlying observations for all three crypto-currencies under consideration.

Since the assumption of non-Gaussian errors would render typical estimation methods like the Kalman filter infeasible, we follow \cite{harrison1976bayesian, west1987scale, gordon1990modeling} and use a scale mixture of Gaussians to approximate the t-distribution,
\begin{align}
\eta_{it}|h_{it} &\sim {t}_{v_i}(0, e^{h_{it}})\quad \Leftrightarrow\quad \eta_{it}|h_{it}, \phi_{it} \sim \mathcal{N}(0, \phi_{it} e^{h_{it}}),\\
 \phi_{it}|v_i &\sim \mathcal{G}^{-1}(v_i/2, v_i/2).
\end{align}
Notice that the degree of freedom parameter $v_i$ is equation-specific, implying that the excess kurtosis of the underlying error distribution is allowed to change across equations, a feature that might be important given the different time series involved. The latent process $\phi_{it}$ simply serves to rescale the Gaussian distribution in case of large shocks. Notice that if $\phi_{it}=1$ for all $i, t$ we obtain the standard time-varying parameter VAR as in \cite{primiceri2005time}.

\subsection{Prior specification}
The prior setup adopted closely follows \cite{feldkircher2017sophisticated}.  More specifically, we use a Normal-Gamma (NG) shrinkage prior on the elements of $\bm{\beta}_{i0}$ and $\sqrt{\bm{\Omega}_i}$.

The NG prior comprises of a Gaussian prior on the coefficients alongside a set of local and global shrinkage parameters for the first $mp$ elements of $\bm{\beta}_{i0}$ and $\text{diag}(\sqrt{\bm{\Omega}_i})$,
\begin{align}
\beta_{ij, 0}|\tau^2_{\beta, ij} &\sim \mathcal{N}(0, \tau^2_{\beta, ij}), \label{eq: priorbeta}\\
\sqrt{\vartheta_{ij}}|\tau^2_{\vartheta, ij} &\sim \mathcal{N}(0, \tau^2_{\vartheta, ij}) \label{eq: priortheta},
\end{align}
for $i=1,\dots, m$ and $j=1,\dots, mp$. Here we let $\tau^2_{s, ij}$  (for  $s \in \{\beta, \vartheta\}$) denote local shrinkage parameters with
\begin{equation}
\tau^2_{s, ij}|\lambda_L \sim \mathcal{G}\left(\kappa, \frac{\kappa \lambda_L}{2}\right).
\end{equation}
 $\kappa$ is a hyperparameter specified by the researcher and $\lambda_L$ is a global shrinkage parameter that is lag-specific, i.e. applied to the elements in $\bm{\beta}_{i0}$ and $\sqrt{\bm{\Omega}_i}$ associated with the $L$th lag of $\bm{y}_t$, and constructed as follows
\begin{equation}
\lambda_L = \prod_{l=1}^L \pi_l, \quad \pi_l \sim \mathcal{G}(c_0, d_0).
\end{equation}
This implies that if $\pi_l>1$, the prior introduces more shrinkage with increasing lag orders. The degree of overall shrinkage is controlled through the hyperparameters $c_0$ and $d_0$.

Notice that this specification pools the parameters that control the amount of time-variation as well as the time-invariant regression parameters. This captures the notion that if a variable is not included initially, the probability of having a time-varying coefficient also decreases (by increasing the lag-specific shrinkage parameter $\lambda_L$).

For the covariance parameters indexed by $j=mp+1,\dots, k_i$ the prior is specified analogously to Eqs. (\ref{eq: priorbeta}) - (\ref{eq: priortheta}) but with $\lambda_L$ replaced by $\varrho$. This choice implies that all covariance parameters as well as the corresponding process innovation variances are pushed to zero simultaneously. For $\varrho$ we again use a Gamma distributed prior,
\begin{equation}
\varrho \sim \mathcal{G}(a_0, b_0),
\end{equation}
with $a_0, b_0$ being hyperparameters.

This prior specification has the convenient property that the parameters $\lambda_L$ and $\varrho$ introduce prior dependence, pooling information across different coefficient types (i.e. regression coefficients and process innovation variances), introducing strong \textit{global} shrinkage on all coefficients concerned. By contrast, the introduction of the local scaling parameters $\tau_{s, ij}$ serves to provide flexibility in the presence of strong overall shrinkage introduced by $\lambda_L$ and $\varrho$. Thus, even if the aforementioned global scaling parameters are large (i.e. heavy shrinkage is introduced in the model), the local scalings provide sufficient flexibility to drag away posterior mass from zero and allowing for non-zero coefficients. The role of the hyperparameter $\kappa$ is to control the tail behavior of the prior. If $\kappa$ is small (close to zero), the prior places more mass on zero but the tails of the marginal prior obtained after integrating over the local scales become thicker \citep[see][for a discussion]{griffin2010inference}.

For the parameters of the log-volatility equation in \autoref{eq:logvola} we follow \cite{kastner2014ancillarity, kastner2015dealing} and use a normally distributed prior on $\mu_j \sim \mathcal{N}(0, 10^2)$, a Beta prior on $\frac{\rho_j +1}{2} \sim \mathcal{B}(25, 5)$ and a Gamma prior on $\varsigma_j \sim \mathcal{G}(1/2, 1/2)$. In addition, we specify a uniform prior on $v_i \sim \mathcal{U}(2, 20)$, effectively ruling out the limiting case of a Gaussian distribution if $v_i$ becomes excessively large.

\subsection{Full conditional posterior simulation}
Estimation of the model is carried out using Markov chain Monte Carlo (MCMC) techniques. Our MCMC algorithm consists of the following blocks:
\begin{enumerate}
\item Conditional on the remaining parameters/states in the model, simulate the full history of $\{\tilde{\bm{\beta}_{it}}\}_{t=1}^T$ using a forward-filtering backward sampling algorithm \citep{carter1994gibbs, fruhwirth1994data} on an equation-by-equation basis.

\item The full history of the log-volatility process as well as the parameters of \autoref{eq:logvola} are obtained by relying on the algorithm proposed in \cite{kastner2014ancillarity} and implemented in the R package stochvol \citep{kastnerjss}.

\item The time-invariant components $\bm{\beta}_{i0}$ as well as $\bm{\theta}_i = \text{diag}(\bm{\Theta}_i)$ are simulated from a multivariate Gaussian posterior that takes a standard form \citep[see][]{feldkircher2017sophisticated}.

\item The sequence of local scaling parameters is simulated from a generalized inverted Gaussian (GIG) distributed posterior distribution given by,
\begin{align}
\tau_{\beta, ij}|\bullet \sim \mathcal{GIG}(\kappa-1/2, \beta_{ij,0}^2, \kappa \lambda_L),   \\
\tau_{\vartheta, ij}|\bullet \sim \mathcal{GIG}(\kappa-1/2, \vartheta_{ij,0}^2, \kappa \lambda_L)
\end{align}
for $j \in \mathcal{A}_L$. The posterior distribution for the scalings associated with the covariance parameters is similar with $\lambda_L$ replaced by $\varrho$.

\item We obtain draws from the posterior of the lag-specific shrinkage parameter associated with the $l$th lag by combining the likelihood $\prod_{i=1}^m \prod_{j \in \mathcal{A}_l} p(\tau^2_{\beta, ij}, \tau^2_{\vartheta, ij}|\pi_l, \lambda_{l-1})$ with the prior on $\pi_l$. The resulting posterior distribution is a Gamma distribution,
\begin{equation}
\pi_l | \bullet \sim \mathcal{G}\left(c_0+\kappa R, d_0 + \lambda_{l-1} \frac{\kappa}{2} \sum_{i=1}^m \sum_{j \in \mathcal{A}_l}(\tau^2_{\beta, ij} + \tau^2_{\vartheta, ij})\right),
\end{equation}
with the $\bullet$ indicating the conditioning on everything else,  $R=2p m^2$ and $\lambda_0 = 1$. The set $\mathcal{A}_l$ selects all coefficients associated with the $l$th lag of $\bm{y}_t$.

Similarly, the conditional posterior of $\varrho$ is given by
\begin{equation}
\varrho|\bullet \sim \mathcal{G}\left(a_0+\kappa\nu, b_0 + \frac{\kappa}{2} \sum_{i=1}^m \sum_{j=mp+1}^{k_i} (\tau^2_{\beta, ij}+\tau^2_{\vartheta, ij})\right),
\end{equation}
where $\nu = m (m-1)$ denotes the number of covariance parameters in addition to the number of process variances for the corresponding parameters.

\item The full history of $\{ \phi_{it} \}_{t=1}^T$ is obtained by independently simulating from an inverted Gamma distribution \citep[see][]{kastner2015heavy},
\begin{equation}
 \phi_{it}| \bullet \sim \mathcal{G}^{-1}\left(\frac{v_i +1}{2}, \frac{v_i + \eta^2_{it} e^{-h_{it}}}{2}\right),
\end{equation}
for $t=1,\dots,T$.

\item To simulate the degrees of freedoms $v_i$, we perform an independent Metropolis Hastings (MH) step described in \cite{kastner2015heavy}.
\end{enumerate}
This algorithm is repeated a large number of times with the first $N^{burn}$ observations being discarded as burn-in.\footnote{In the empirical application we use 30,000 overall iterations with the first 15,000 being discared as burn-in.} Notice that the equation-by-equation algorithm yields significant computational gains relative to competing estimation algorithms that rely on full-system estimation of the VAR model.

\section{Forecasting results}\label{sec:forecasting}
\subsection{Model specification and design of the forecasting exercise}
In this section, we briefly describe model specification and the design of the forecasting exercise. The prior setup for our benchmark specification (henceforth labeled the t-TVP NG) model closely follows the existing literature on NG shrinkage priors  \citep{griffin2010inference, bitto2016achieving, huber2017adaptive, feldkircher2017sophisticated}. More specifically, we set $\kappa=0.1$, $c_0=1.5$ and $c_1=1$ to center the prior on $\pi_l$ above unity while $a_0=b_0=0.01$.  The choice for $\kappa$ implies that we place a large amount of prior mass on zero while at the same time allow for relatively thick tails. Our choice for the Gamma prior on $\varrho$ introduces heavy shrinkage on the covariance parameters as well as the corresponding process standard deviations.

For all models (i.e. the competitors introduced in the next subsection) we consider as well as the proposed model we include a single lag of the endogenous variables. Higher lag orders are generally possible but given the high dimension of the state space and the increased computational complexity we stick to one lag. In addition, experimenting with slightly higher lag orders leads to models that are relatively unstable during several points in time in our estimation sample.

The design of our forecasting exercise is the following. We start with an initial estimation period that spans the period between the end of November 2016 (22nd of November)  to the end  of April 2017 (26th of April ). The remaining 160 days are used as a hold-out period. After obtaining the one-step-ahead predictive density for the 27th of April 2017, we consequently expand the estimation sample by a single day until the end of the sample is reached. This yields a sequence of 160 one-day-ahead predictive densities.

To assess the predictive fit of our model we use the log-predictive likelihood (LPL), motivated in, e.g., \cite{geweke2010comparing}, and the root mean square forecast error (RMSE). Using LPLs enables us to assess not only how well the model fits in terms of point predictions but also how well higher moments of the predictive density are captured.  In addition, to assess model calibration we use univariate probability integral transforms \citep{diebold1998evaluating, clark2011real, amisano2017prediction}.

\subsection{Competing models}
Our set of competing models ranges from univariate benchmark models that feature SV to a wide set of multivariate benchmark models.  The first set of models considered are  a  random walk (RW-SV) and the AR(1) model (henceforth labeld AR-SV), both estimated with SV. We use non-informative priors on the AR(1) regression coefficient and the same prior setup for the log-volatility equation as discussed in the previous section.  These two models serve to illustrate whether a multivariate modeling approach pays off and, in addition, whether allowing for structural changes in the underlying regression parameters improves predictive capabilities.

In addition, we consider a set of nested multivariate benchmark models. To quantify the accuracy gains of time-varying parameter specifications, we estimate  three constant parameter VARs with SV. The first VAR uses the prior setup described above but with  $\sqrt{\vartheta}_{ij}=0$ for all $i, j$. The second model is a non-conjugate Minnesota VAR with asymmetric shrinkage across equations. To select the hyperparameters we follow \cite{giannone2015prior} and place hyperpriors on all hyperparameters and estimate them using a random walk Metropolis Hastings step. The last VAR we consider is a model that features a stochastic search variable selection (SSVS) prior specified as  in \cite{george2008bayesian}. This implies that a two component Gaussian prior is used with the Gaussians differing in terms of their prior variance. One component features a large prior variance (labeled the slab distribution) which introduces relatively little prior information whereas the second component has a prior variance close to zero (the spike component) that strongly forces the posterior of the respective coefficient to zero. We set the hyperparameters (i.e. the prior standard deviations) for the slab distribution by using the OLS standard deviation times a constant (ten in our case) while the prior standard deviation on the spike component is set equal to $0.1$ times the OLS standard deviation.

Moreover, we include two time-varying parameter models with SV and Gaussian measurement errors. The first TVP-VAR considered (labeled TVP) is based on  an uninformative prior (obtained by setting the prior variances to unity for both, the initial states as well as the process standard deviations). The next benchmark model (called TVP NG) is our proposed specification with a NG prior but with Gaussian errors (i.e. $\phi_{it}=1$ for all $i,t$). This choice serves to assess whether additional flexibility on the measurement errors is needed.

Finally, the last model considered is the most flexible specification in terms of the law of motion of the latent states. This model, labeled the threshold TVP-VAR (labeled TTVP) is based on \cite{huber2017new} and captures the notion that parameter movements are only allowed if they are sufficiently large. To achieve this, a threshold specification for the process variances is adopted. This specification depends on a latent indicator that, in turn, is driven by the absolute size of parameter changes. Thus, if the change in a given regression parameter is large (i.e. exceeds a certain threshold we estimate), we use a large variance in \autoref{eq: states}. By contrast, if the change is small the process variance is set to a small constant that is close to zero. The prior specification adopted here closely follows the benchmark specification outlined in \cite{huber2017new} and we refer to the original paper for additional details.


\subsection{Out of sample forecasting performance}
We start by considering the forecasting performance in terms of log predictive likelihoods (LPS). \autoref{tab: LPS_full} displays the LPS as well as the RMSEs for the competing models. The first column shows the joint LPS for the three crypto-currencies considered while the next three columns display the marginal LPS for a given crypto-currency. The final three columns show the RMSEs.

\begin{table}[t]
\centering
\begin{tabular}{lrrrrrrrrr}
  \toprule
  & & \multicolumn{3}{c}{Log predictive score} & & \multicolumn{3}{c}{Root mean square error} \\
 & JointLPS &  & Bitcoin & Litecoin & Ethereum &  & Bitcoin & Litecoin & Ethereum \\
  \midrule
TTVP & 621.023 &  & 286.360 & 134.231 & 153.201 &  & 0.050 & 0.084 & 0.078 \\
  TVP & 451.631 &  & 187.474 & 106.946 & 97.300 &  & 0.074 & 0.133 & 0.134 \\
  TVP NG & 632.410 &  & 286.134 & 144.629 & 159.562 &  & 0.050 & 0.083 & 0.079 \\
  t-TVP NG & 643.873 &  & 277.679 & 161.768 & 166.988 &  & 0.050 & 0.084 & 0.078 \\ \midrule
  Minn-VAR & 577.779 &  & 283.399 & 123.580 & 153.274 &  & 0.051 & 0.085 & 0.078 \\
  NG-VAR & 592.391 &  & 286.483 & 130.194 & 148.553 &  & 0.051 & 0.084 & 0.078 \\
  SSVS & 586.083 &  & 286.255 & 122.346 & 153.081 &  & 0.051 & 0.084 & 0.078 \\ \midrule
  RW-SV & 483.952 &  & 240.751 & 131.410 & 112.487 &  & 0.073 & 0.112 & 0.114 \\
  AR-SV & 598.936 &  & 280.487 & 158.899 & 159.725 &  & 0.051 & 0.085 & 0.078 \\
   \bottomrule
\end{tabular}\caption{Joint and marginal log predictive likelihoods for all models considered (left panel) and root mean square forecast errors (right panel). For the joint log predictive likelihood we integrate out the effect of the other variables included in $\bm{y}_t$ and focus exclusively on the predictive performance for the three crypto-currencies.}\label{tab: LPS_full}
\end{table}

Considering the joint LPS indicates that across models, the t-TVP NG specification outperforms the remaining models. This points towards the necessity to  allow for both,  a flexible error distribution as well as  time-varying parameters with appropriate shrinkage priors. Especially when compared to the constant parameter VAR models, all three TVP-VAR specifications with some form of shrinkage yield pronounced accuracy gains. Notice also that the AR(1) model with SV proves to be a tough competitor relative to the set of Bayesian VARs.

The necessity of introducing shrinkage in the TVP-VAR framework can be seen by comparing the joint forecasting performance of the TVP model with the remaining TVP-VARs considered. Notice  that in our medium-scale model, a TVP-VAR with relatively little shrinkage leads to overfitting issues which in turn are detrimental for forecasting performance.

Zooming into the results for the three crypto-currencies, we generally observe that models performing well in terms of the joint LPS also do well on average. One interesting exception is our proposed t-TVP NG specification. While the performance gains for Litecoin and Ethereum appear to be substantial vis-a-vis the competing models, we find that Bitcoin predictions appear to be inferior relative to the TTVP and the TVP NG specifications. If the researcher is interested in predicting the price of Bitcoin, the two best performing models are the TTVP specification and the Bayesian VAR with a Normal-Gamma shrinkage prior. Interestingly, notice that the comparatively weaker joint performance of the BVAR models stems from weaker Litecoin and Ethereum predictions whereas Bitcoin predictions appear to be rather precise.

Considering point forecasting performance generally corroborates the findings for density forecasts. Here we again observe that models which yield precise predictive densities also work well when only point predictions are  considered. Notice, however, that the differences in terms of RMSE between multivariate models and the univariate AR(1) model are negligible. This somewhat highlights that forecasting gains in terms of predictive likelihoods stem from higher moments of the predictive density like the predictive variance (in terms of the marginal log scores) or a more appropriate modeling strategy for the predictive variance-covariance structure.

\begin{figure}[t]
\begin{minipage}[]{.45\linewidth}
\subcaption{Bitcoin}
\centering
\includegraphics[scale=.3, trim=20 35 10 40, clip]{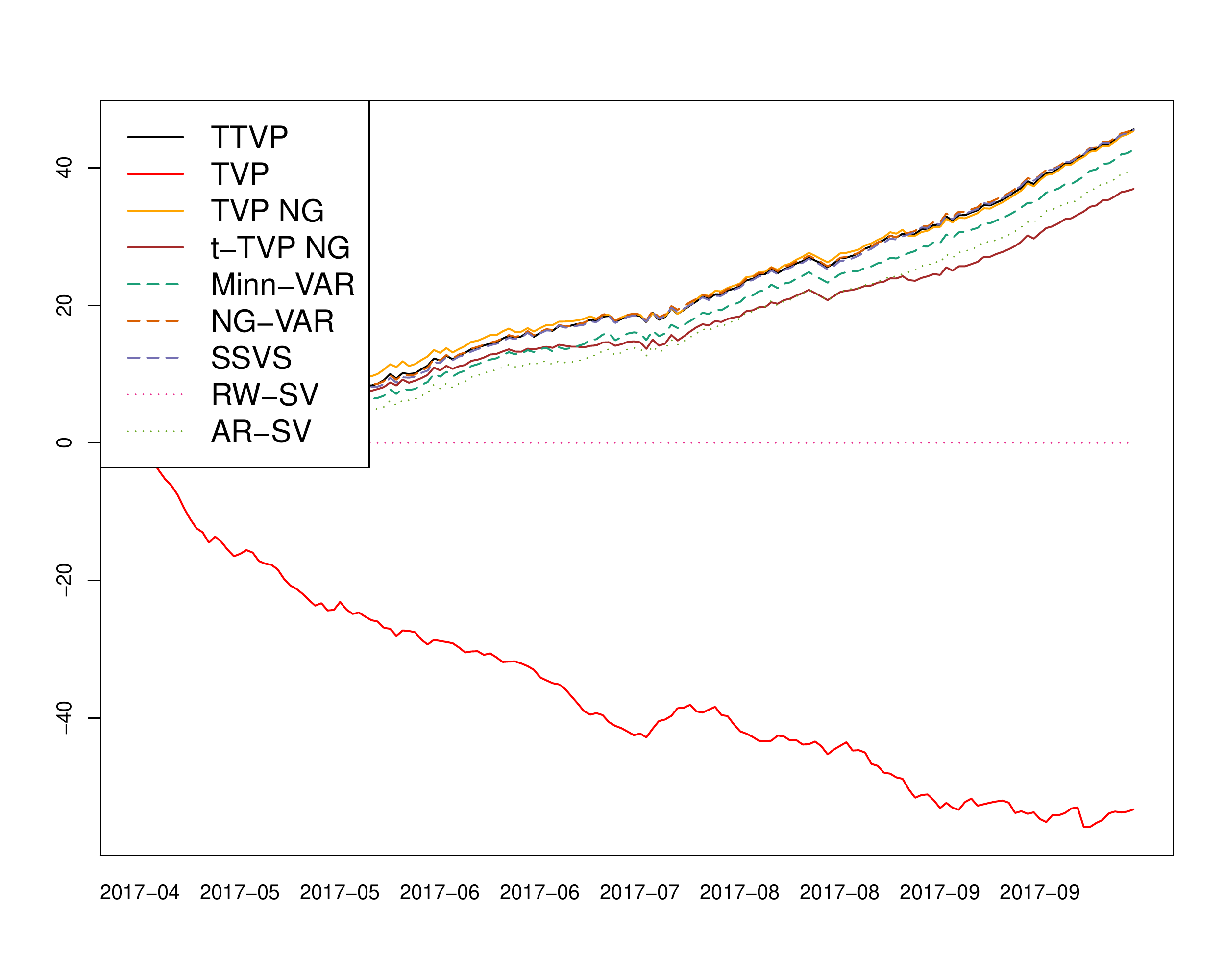}
\end{minipage}
\begin{minipage}[]{.45\linewidth}
\subcaption{Litecoin}
\centering
\includegraphics[scale=.3, trim=20 35 10 40, clip]{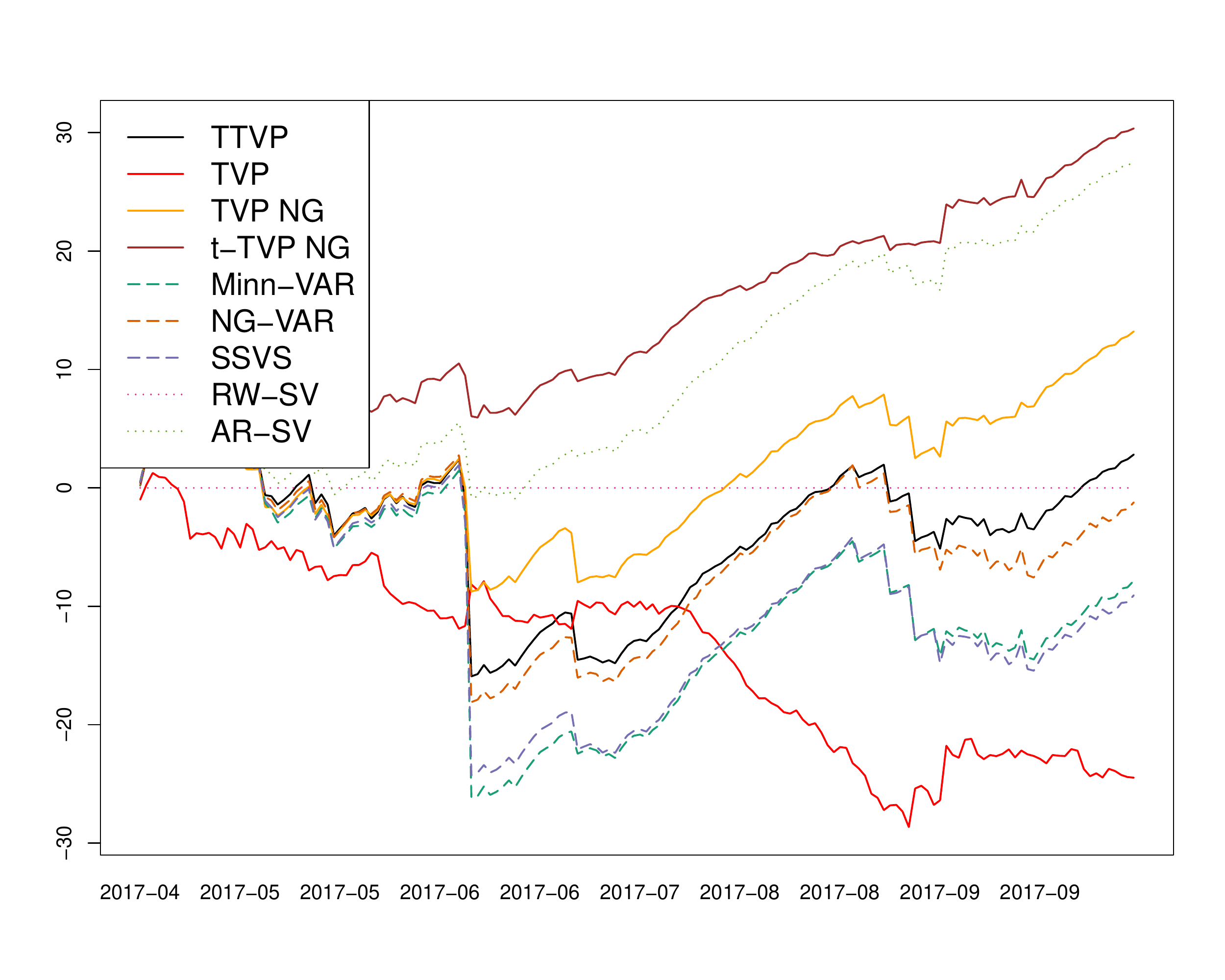}
\end{minipage}\\
\begin{minipage}[]{.45\linewidth}
\subcaption{Ethereum}
\centering
\includegraphics[scale=.3, trim=20 35 10 40, clip]{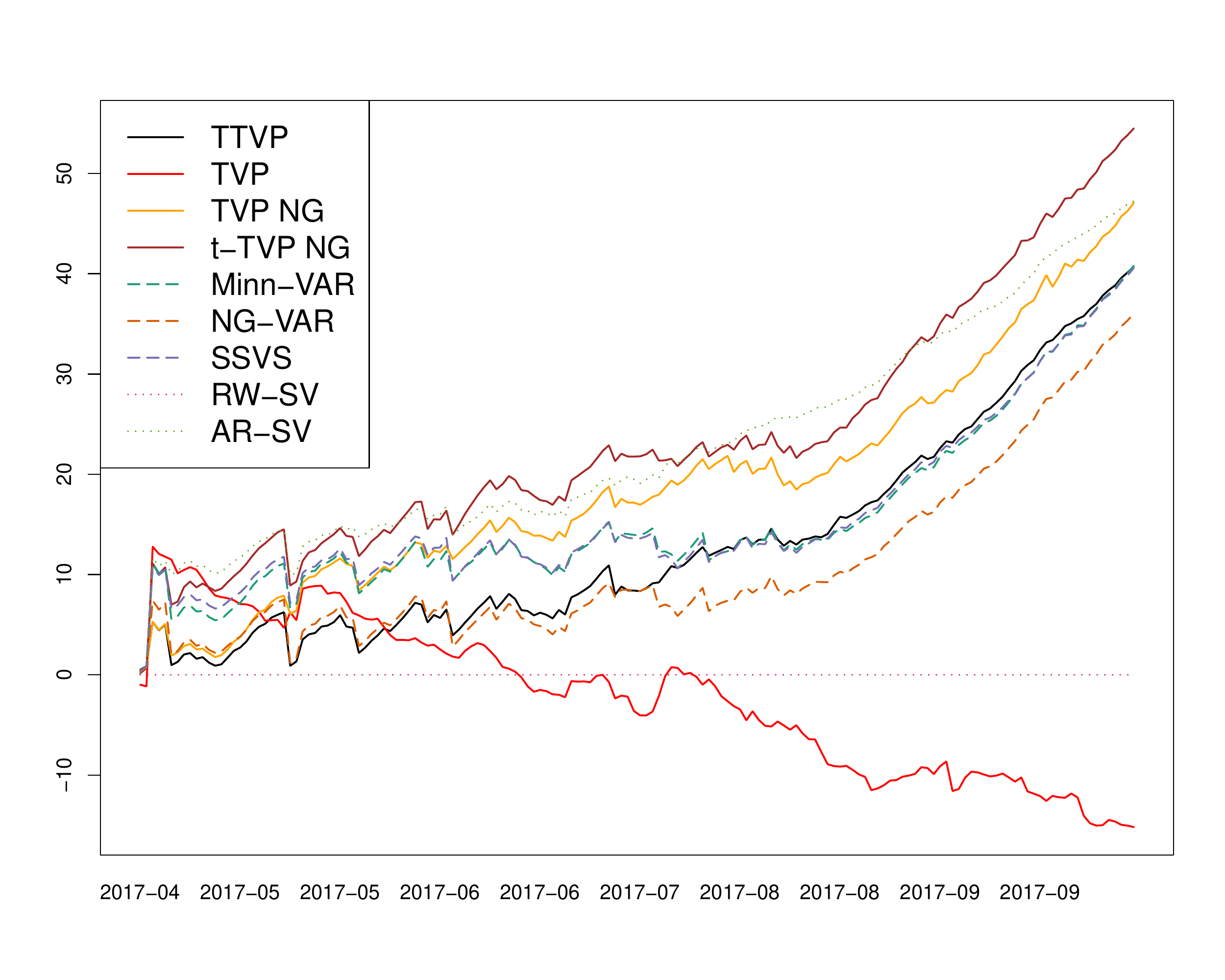}
\end{minipage}
\begin{minipage}[]{.45\linewidth}
\subcaption{Log predictive likelihood}
\centering
\includegraphics[scale=.3, trim=20 35 10 40, clip]{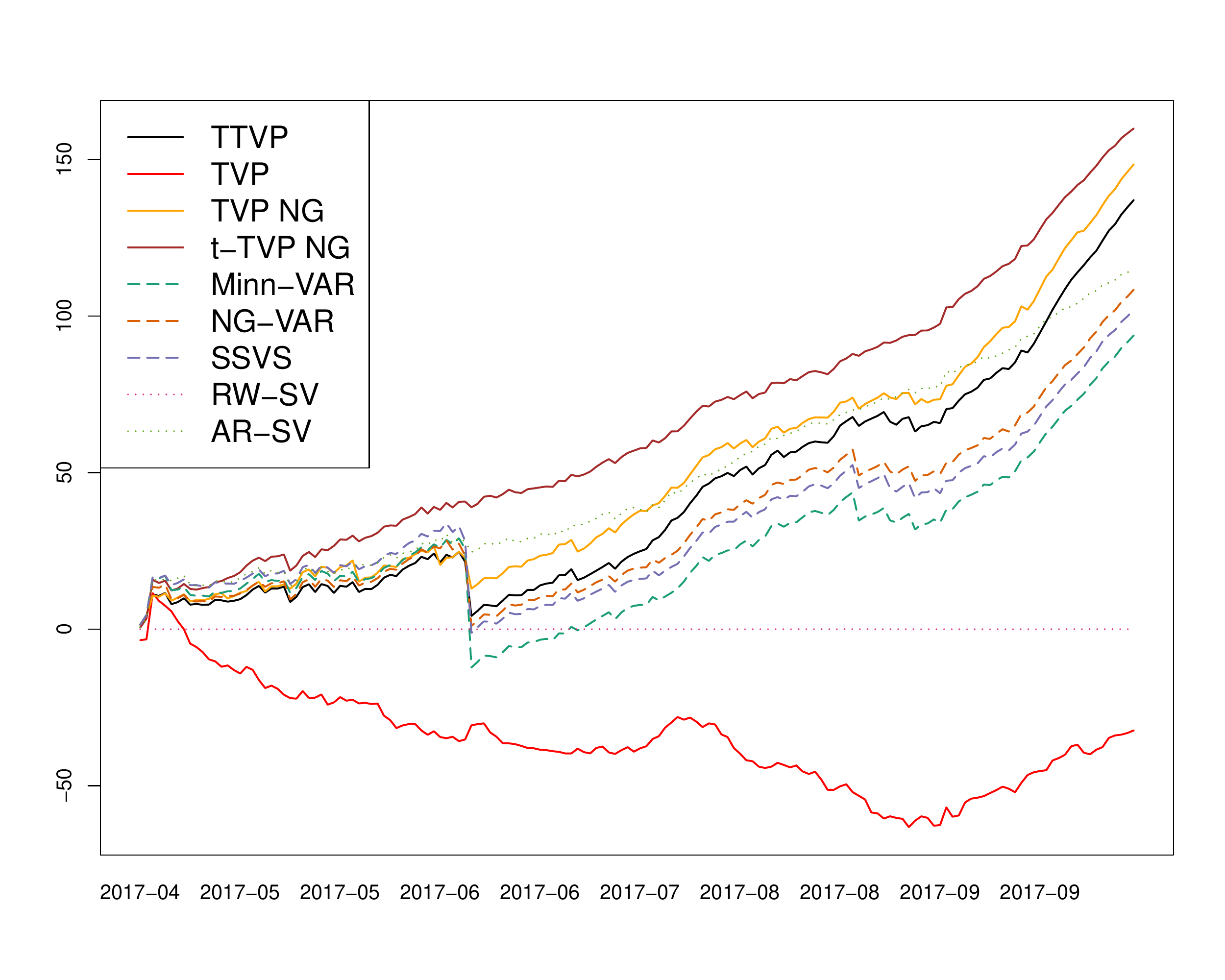}
\end{minipage}\\
\caption{Log predictive Bayes factors relative to the TVP-VAR over time.}\label{fig:LPSovertime}
\end{figure}

Next, we investigate whether differences in forecasting performance appear to be time-varying. \autoref{fig:LPSovertime} shows the log predictive Bayes factors relative to the random walk with SV.  Comparing the model performances over time points towards a pronounced degree of heterogeneity over time. For Bitcoin (see panel (a)) shows that the two best performing models are the TTVP and the TVP NG specifications. While the former yields a slightly better performance over time, the latter proves to be the best performing model during the first part of the hold-out period. For the remaining models we find only relatively little time-variation in their predictive performance.
Considering the results for Litecoin (see panel (b)) we find pronounced movements in relative forecasting accuracy.  More specifically, we find that while forecasting performance appears to be homogeneous during the first months of the hold-out period. From May 2017 onward, the t-TVP NG specification starts to perform extraordinarily well, improving upon all competitors by large margins. 

Finally, panels (c) and (d) show the performance for Ethereum as well as the overall performance over time. Here we generally find results that are comparable with the findings described above. Notice that the overall log predictive likelihood displays a pattern similar to the one of the marginal LPS for the remaining crypto-currencies. However, compared to panel (a) we observe that the t-TVP specification also excels in terms of joint density predictions. The main difference is that the superior performance of the t-TVP NG model in terms of predicting Litecoin prices lifts the log predictive Bayes factor above the ones obtained for all competing models.

\subsection{Model evaluation using probability integral transforms}
Following \cite{diebold1998evaluating, clark2011real, amisano2017prediction}, if a given model  $\mathcal{M}_i$ is correctly specified one can show that
\begin{equation}
z_{jt, i} = \Phi^{-1}(F_y({y}_{jt}| \boldsymbol{y}_{1:t-1}, \mathcal{M}_i)) \stackrel{iid}{\sim} \mathcal{N}(0,1),
\end{equation}
for $t=t_0, \dots, T$ and $j=1,\dots, m$ and $t_0$ indicating the first observation of the hold-out period (i.e. 22nd of November). Hereby we let $\Phi^{-1}$ denote the inverse distribution function of the standard normal distribution and $F_y({y}_{jt}| \boldsymbol{y}_{1:t-1}, \mathcal{M}_i)$ denotes the cumulative distribution function associated with the underlying predictive distribution of model $i$. If the model is correctly specified the sequence of normalized forecast errors $\{z_{jt}\}_{t=t_0}^T$ is independent and identically standard normally distributed.

\begin{figure}[h!]
\centering
\begin{minipage}[]{1\linewidth}
\subcaption{Bitcoin}
\centering
\includegraphics[scale=.55]{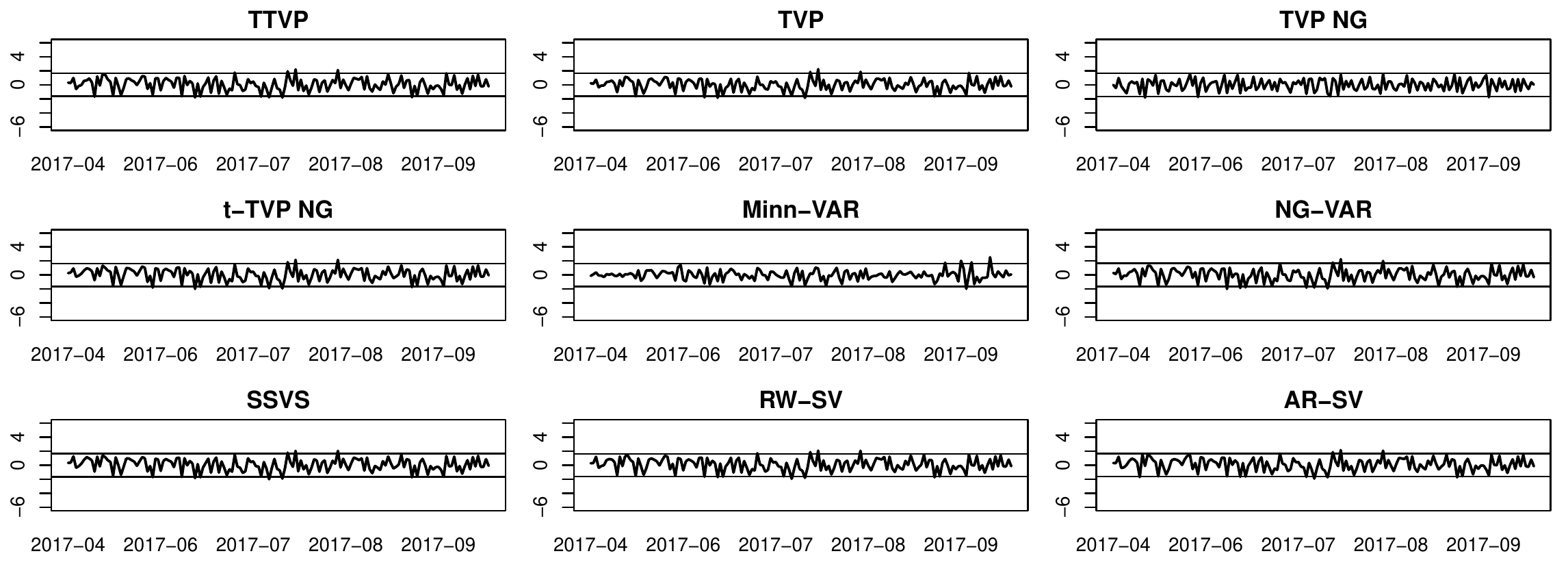}
\end{minipage}
\begin{minipage}[]{1\linewidth}
\subcaption{Litecoin}
\centering
\includegraphics[scale=.55]{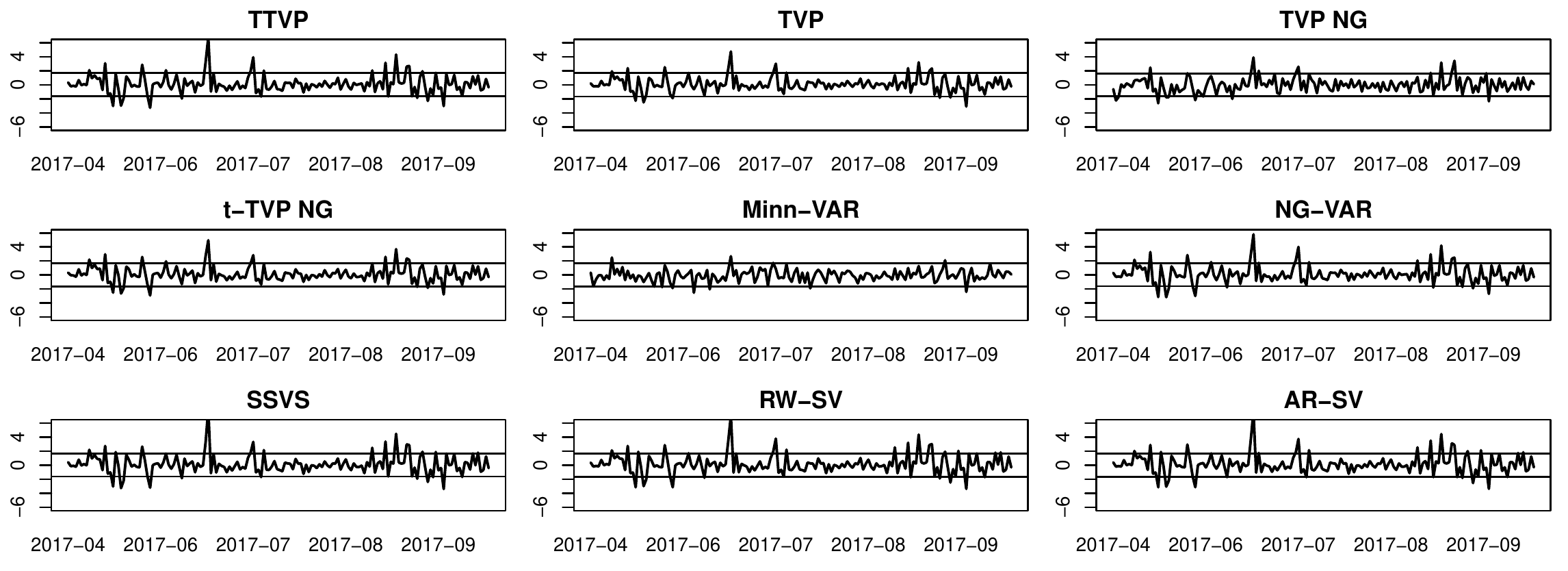}
\end{minipage}\\
\begin{minipage}[]{1\linewidth}
\subcaption{Ethereum}
\centering
\includegraphics[scale=.55]{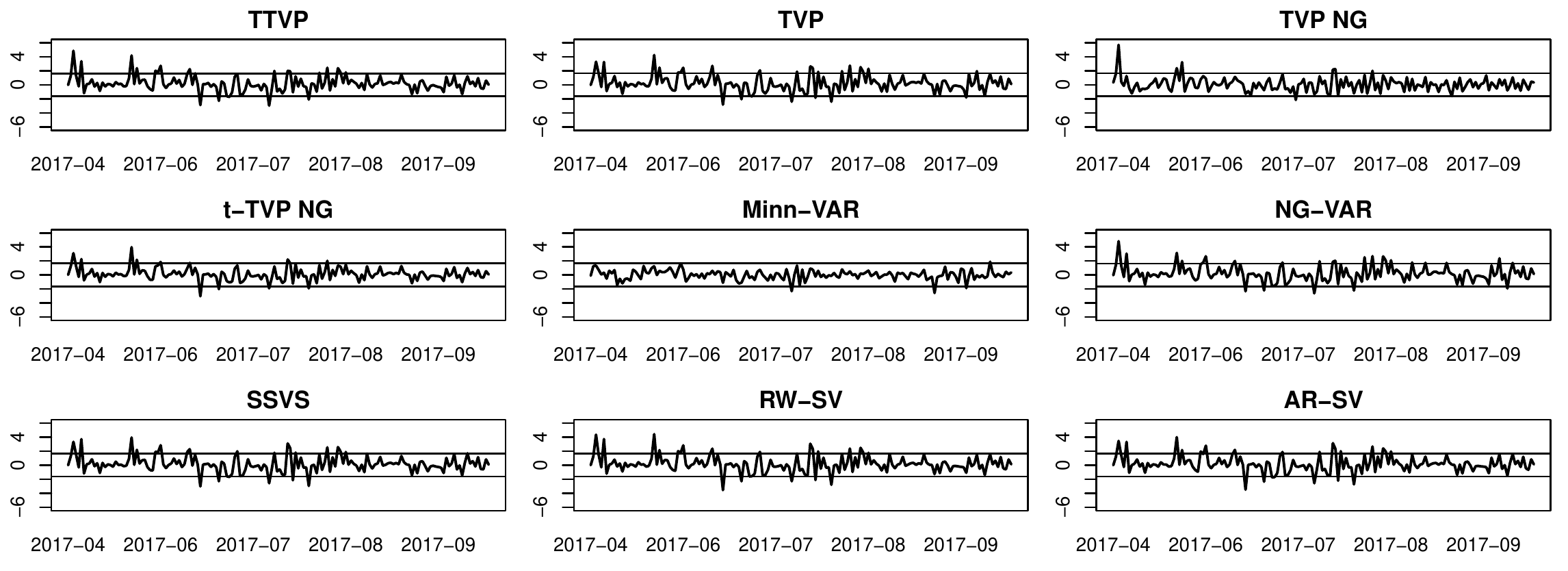}
\end{minipage}
\caption{Normalized forecast errors across models and crypto-currencies. Errors are obtained by applying the inverse cumulative distribution function of a normal distribution to the PIT of the one-step-ahead forecast errors.}\label{fig:normerrors}
\end{figure}

\autoref{fig:normerrors} (a) to (c) shows the normalized forecast errors across models and for all three crypto-currencies considered while \autoref{tab:pit_results} provides statistical tests that aim to support our visual assessment of \autoref{fig:normerrors}. In the case of Bitcoin and Litecoin, we find that the mean appears to be close to zero. This finding is corroborated by the first column in \autoref{tab:pit_results} which displays the empirical mean obtained by regressing $z_{jt, i}$ on a constant, with $p$ values in parentheses. Notice that for Ethereum, we find the normalized forecast errors of the majority of models to be centered above zero. The two exceptions are the TVP NG specification and the Minnesota prior VAR. Considering again panel (c) reveals that these deviations from zero are mainly driven by the failure to capture the conditional mean during the beginning of the hold-out period.

\begin{table}[h!]
\centering
\scalebox{0.9}{
\begin{tabular}{lrrr}
  \toprule
 & Mean (p-value) & Variance (p-value) & Persistence (p-value) \\
  \midrule
  \multicolumn{4}{c}{Bitcoin}\\
TTVP1 & 0.060 (0.401) & 0.821 (0.024) & -0.078 (0.329) \\
  TVP1 & 0.013 (0.838) & 0.649 (0.000) & -0.085 (0.283) \\
  TVP NG1 & 0.004 (0.948) & 0.683 (0.000) & -0.439 (0.000) \\
  t-TVP NG1 & 0.051 (0.466) & 0.783 (0.005) & -0.060 (0.454) \\
  Minn-VAR1 & 0.007 (0.902) & 0.490 (0.000) & -0.135 (0.089) \\
  NG-VAR1 & 0.022 (0.756) & 0.809 (0.018) & -0.093 (0.243) \\
  SSVS1 & 0.058 (0.410) & 0.799 (0.007) & -0.052 (0.513) \\
  RW-SV1 & 0.082 (0.246) & 0.796 (0.007) & -0.058 (0.470) \\
  AR-SV1 & 0.098 (0.168) & 0.804 (0.011) & -0.051 (0.518) \\
    \multicolumn{4}{c}{Litecoin}\\
  TTVP2 & 0.121 (0.255) & 1.790 (0.030) & 0.023 (0.772) \\
  TVP2 & 0.096 (0.254) & 1.120 (0.544) & 0.011 (0.891) \\
  TVP NG2 & -0.009 (0.912) & 1.154 (0.347) & -0.052 (0.516) \\
  t-TVP NG2 & 0.115 (0.202) & 1.295 (0.187) & 0.027 (0.731) \\
  Minn-VAR2 & -0.049 (0.472) & 0.732 (0.007) & -0.084 (0.292) \\
  NG-VAR2 & 0.114 (0.254) & 1.596 (0.047) & 0.008 (0.917) \\
  SSVS2 & 0.128 (0.253) & 2.001 (0.031) & 0.018 (0.821) \\
  RW-SV2 & 0.144 (0.188) & 1.920 (0.018) & 0.017 (0.831) \\
  AR-SV2 & 0.152 (0.177) & 2.020 (0.021) & 0.025 (0.756) \\
    \multicolumn{4}{c}{Ethereum}\\
  TTVP3 & 0.201 (0.025) & 1.285 (0.212) & 0.121 (0.127) \\
  TVP3 & 0.208 (0.026) & 1.393 (0.047) & 0.059 (0.461) \\
  TVP NG3 & 0.090 (0.250) & 0.980 (0.93) & -0.026 (0.743) \\
  t-TVP NG3 & 0.148 (0.042) & 0.848 (0.306) & 0.072 (0.367) \\
  Minn-VAR3 & 0.023 (0.672) & 0.478 (0.000) & 0.047 (0.556) \\
  NG-VAR3 & 0.223 (0.014) & 1.335 (0.107) & 0.100 (0.207) \\
  SSVS3 & 0.188 (0.043) & 1.393 (0.056) & 0.075 (0.343) \\
  RW-SV3 & 0.194 (0.040) & 1.429 (0.071) & 0.065 (0.417) \\
  AR-SV3 & 0.176 (0.058) & 1.380 (0.065) & 0.052 (0.514) \\
   \bottomrule
\end{tabular}
}
\caption{Statistical results for the transformed forecast errors}\label{tab:pit_results}
\end{table}

Considering the variances reveals that in the case of Bitcoin, the variances of the normalized errors are all well below unity, indicating that the estimated predictive variance is generally too high.  Put differently, this is an indication for a situation where too many actual observations fall in the center of the predictive distribution. This finding appears to be strongly supported by the second column of \autoref{tab:pit_results}, which displays the estimated variance of the normalized forecast error obtained by regressing the squared error on a constant. For the t-TVP NG and TTVP specifications we find slightly higher variances.  Our interpretation is that allowing for a flexible error specification either by directly using non-Gaussian shocks in conjunction with stochastic volatility or by introducing more flexibility on the law of motion of the latent states slightly helps to push the variances towards one.

For Litecoin and Ethereum, the variances appear to be closer to one for all TVP specifications except for the TTVP model (in the case of Litecoin). It is noteworthy that especially for Litecoin, constant parameter models with SV tend to either underestimate the predictive variance or fail to capture observations in the tail of the empirical distribution.

Finally, considering the persistence of $z_{jt, i}$ reveals that most models tend to produce normalized errors which display muted persistence levels. This is corroborated by the final column of \autoref{tab:pit_results} which shows the persistence parameter obtained by estimating AR(1) models in $z_{jt, i}$ along with its $p$-values.

\section{Economic performance criteria: A simple trading exercise}\label{sec:trading}
To assess which model performs well in terms of economic performance criteria, we perform a trading exercise where each model is used to generate a set of optimal weights attached to each of the three crypto-currencies considered. Using the models discussed in the previous sections as well as two additional investment strategies that are based on  equal weights and a simple passive investments in Bitcoin allows us to infer whether constructing a trading strategy based on more sophisticated econometric models pays off in terms of generating superior returns.

We assume that investors adopt two strategies to find a optimal sequence of weights $\boldsymbol{w}_{it}=(w_{1i, t}, w_{2i, t}, w_{3i, t})'$. The first one is the standard minimum variance portfolio problem  that aims to allocate money between the three assets considered such that the portfolio variance is minimized. This implies that for $t = t_0, \ldots, T$ the investor solves
\begin{equation}
\begin{aligned}
& \underset{\boldsymbol{w}_{it}}{\text{minimize}}
& & \boldsymbol{w}_{it}\bm{P}_{i, t|t-1} \boldsymbol{w}'_{it} \\
& \text{subject to}
& & \bm{1}'\boldsymbol{w}_{it}=1,
\end{aligned}\label{eq:minVAR}
\end{equation}
where $\bm{1}$ is a $3$-dimensional vector of ones and $\bm{P}_{i, t|t-1}$ denotes the variance of model $i$'s one-step-ahead predictive density.

The second strategy adds a specific portfolio target return to the optimization problem in  \autoref{eq:minVAR}, i.e.,
\begin{equation}
\bm{w}'_{it} \bm{\mu}_{it|t-1} \ge r^*_t.
\end{equation}
Here we let $\bm{\mu}_{it|t-1}$ denote the one-step-ahead predictive mean of model $i$ and $r^*_t$ is a potentially time-varying target return the investor wants to match. This strategy, called the target mean-variance portfolio, tries to minimize the overall portfolio variance while at the same time maintaining the desired return $r_t^*$ \citep[see][]{markowitz1952}.
\begin{table}[t]
\centering
\begin{tabular}{lrrrr}
  \toprule
   & Min-Variance  & \multicolumn{3}{c}{Target mean-variance} \\
 &  & $r^*=\frac{0.10}{252}$ & $\frac{0.15}{252}$ & $\frac{0.30}{252}$ \\
  \midrule
TTVP & 2.379 & 2.900 & 2.923 & 2.978 \\
  TVP & 2.579 & 2.015 & 2.019 & 2.031 \\
  TVP NG & 2.510 & 2.069 & 2.053 & 1.995 \\
  t-TVP NG & 2.365 & 2.452 & 2.465 & 2.498 \\
  Minn-VAR & 2.066 & -0.313 & -0.243 & 0.004 \\
  NG-VAR & 2.023 & 2.845 & 2.725 & 2.312 \\
  SSVS & 1.997 & 2.942 & 2.948 & 2.943 \\
  RW-SV & 2.040 & 1.399 & 1.415 & 1.464 \\
  AR-SV & 2.201 & 2.390 & 2.407 & 2.453 \\
  Equal weights & 2.528 & 2.528 & 2.528 & 2.528 \\
  only BTC & 2.419 & 2.419 & 2.419 & 2.419 \\
   \bottomrule
\end{tabular}
\caption{Annualized sharpe ratios of various competing investment strategies over the hold-out sample. Min-Variance refers to the minimum variance portfolio whereas target mean-variance is the target mean-variance portfolio for different target returns. Equal weights refers to using $w_{jt}=1/3$ for all $j, t$ and only BTC sets the weight associated with Bitcoin equal to one.}\label{tab:sharperatios}
\end{table}

\autoref{tab:sharperatios} shows annualized Sharpe ratios for the minimum-variance portfolio strategy as well as for the target mean-variance portfolio strategy for $r_t^*=r^* \in \{\frac{0.10}{252},\frac{0.15}{252},\frac{0.30}{252}\}$. Considering the performance of the minimum variance portfolio (see first column in \autoref{tab:sharperatios}) shows that performance differences across models appear to be relatively small. This indicates that weights generated by the set of econometric models are similar, and when compared to the other strategies, more stable over time. Inspection of the weights (not shown) also suggests that this strategy yields weights that are seldom above one  in absolute values (i.e.  leveraged long/short positions). The single best performing model is the no-shrinkage TVP specification, closely followed by the TVP NG model. Notice that using simple equal weights also yields favorable  risk/return ratios.

Considering the target mean-variance strategy for different target returns yields more heterogeneous  model performances. The two best performing models are the TTVP model and the constant parameter VAR coupled with the SSVS prior. For the TVP VAR and the TVP NG model,  we find that performance decreases when compared to the minimum variance portfolio strategy while for the proposed t-TVP NG we observe increasing Sharpe ratios. Comparing different $r^*$ yields no discernible  differences, with most models that do well for modest target returns also performing well if target returns become more ambitious.

Across strategies it is worth noting that performing a passive investment in Bitcoin only (i.e. setting the corresponding weight equal to one for all $t$) also works well but one could still improve upon that strategy by considering more flexible portfolio allocation strategies.
\section{Conclusive remarks}
In this paper we perform a systematic comparison of univariate and multivariate time series models in terms of predicting one-day-ahead returns for three crypto-currencies, namely Bitcoin, Litecoin and Ethereum. To match the pronounced degree of volatility observed in daily returns of crypto-currencies, we propose a medium-scale multivariate state space model that features heavy-tailed measurement errors and stochastic volatility, a feature that turns out to be advantageous for density predictions. More generally, we find that it pays off to allow for time-varying parameters and a flexible error distribution only if suitable shrinkage priors are introduced. These priors introduce shrinkage to select the subset of time-varying coefficients in a flexible manner. To gauge the economic significance of our findings we also perform a trading exercise. The results show that models which perform well in forecasting also tend to work well when used to guide investment decisions.

\normalsize
\singlespacing
\bibliographystyle{./bibtex/fischer}
\bibliography{./bibtex/favar,./bibtex/mpShocks,./bibtex/References,./bibtex/zoerner}

\begin{thebibliography}{37}
\providecommand{\natexlab}[1]{#1}

\bibitem[{Amisano and Geweke(2017)}]{amisano2017prediction}
Amisano G and Geweke J (2017) Prediction using several macroeconomic models.
\newblock \emph{Review of Economics and Statistics} 99(5), 912--925

\bibitem[{Belmonte et~al.(2014)Belmonte, Koop and
  Korobilis}]{belmonte2014hierarchical}
Belmonte MA, Koop G and Korobilis D (2014) Hierarchical shrinkage in
  time-varying parameter models.
\newblock \emph{Journal of Forecasting} 33(1), 80--94

\bibitem[{Bitto and Fr{\"u}hwirth-Schnatter(2016)}]{bitto2016achieving}
Bitto A and Fr{\"u}hwirth-Schnatter S (2016) Achieving shrinkage in a
  time-varying parameter model framework.
\newblock \emph{arXiv preprint arXiv:1611.01310}

\bibitem[{B{\"o}hme et~al.(2015)B{\"o}hme, Christin, Edelman and
  Moore}]{bohme2015bitcoin}
B{\"o}hme R, Christin N, Edelman B and Moore T (2015) Bitcoin: Economics,
  technology, and governance.
\newblock \emph{The Journal of Economic Perspectives} 29(2), 213--238

\bibitem[{Carlin et~al.(1992)Carlin, Polson and Stoffer}]{carlin1992monte}
Carlin BP, Polson NG and Stoffer DS (1992) A Monte Carlo approach to nonnormal
  and nonlinear state-space modeling.
\newblock \emph{Journal of the American Statistical Association} 87(418),
  493--500

\bibitem[{Carriero et~al.(2015)Carriero, Clark and
  Marcellino}]{carriero2015large}
Carriero A, Clark TE and Marcellino M (2015) Large vector autoregressions with
  asymmetric priors.
\newblock Technical report, Working Paper, School of Economics and Finance,
  Queen Mary University of London

\bibitem[{Carter and Kohn(1994)}]{carter1994gibbs}
Carter CK and Kohn R (1994) On Gibbs sampling for state space models.
\newblock \emph{Biometrika} 81(3), 541--553

\bibitem[{Cheah and Fry(2015)}]{cheah2015speculative}
Cheah ET and Fry J (2015) Speculative bubbles in Bitcoin markets? An empirical
  investigation into the fundamental value of Bitcoin.
\newblock \emph{Economics Letters} 130, 32--36

\bibitem[{Chiu et~al.(2017)Chiu, Ching-Wai, Mumtaz, Haroon and
  Pint{\'e}r}]{mumtaz2017forecasting}
Chiu, Ching-Wai, Mumtaz, Haroon and Pint{\'e}r G (2017) Forecasting with VAR
  models: Fat tails and stochastic volatility.
\newblock \emph{International Journal of Forecasting} 33(4), 1124--1143

\bibitem[{Chu et~al.(2017)Chu, Chan, Nadarajah and Osterrieder}]{chu2017garch}
Chu J, Chan S, Nadarajah S and Osterrieder J (2017) GARCH Modelling of
  Cryptocurrencies.
\newblock \emph{Journal of Risk and Financial Management} 10(4), 17

\bibitem[{Clark(2011)}]{clark2011real}
Clark TE (2011) Real-time density forecasts from Bayesian vector
  autoregressions with stochastic volatility.
\newblock \emph{Journal of Business \& Economic Statistics} 29(3)

\bibitem[{Clark and Ravazzolo(2015)}]{clark2015macroeconomic}
Clark TE and Ravazzolo F (2015) Macroeconomic Forecasting Performance under
  Alternative Specifications of Time-Varying Volatility.
\newblock \emph{Journal of Applied Econometrics} 30(4), 551--575

\bibitem[{Cogley and Sargent(2005)}]{CogleySargent2005}
Cogley T and Sargent TJ (2005) {Drift and Volatilities: Monetary Policies and
  Outcomes in the Post WWII U.S}.
\newblock \emph{Review of Economic Dynamics} 8(2), 262--302

\bibitem[{Diebold et~al.(1998)Diebold, Gunther and Tay}]{diebold1998evaluating}
Diebold FX, Gunther TA and Tay AS (1998) Evaluating density forecasts with
  applications to financial risk management.
\newblock \emph{International Economic Review} 39(4), 863

\bibitem[{Feldkircher et~al.(2017)Feldkircher, Huber and
  Kastner}]{feldkircher2017sophisticated}
Feldkircher M, Huber F and Kastner G (2017) Sophisticated and small versus
  simple and sizeable: When does it pay off to introduce drifting coefficients
  in Bayesian VARs?
\newblock \emph{arXiv preprint arXiv:1711.00564}

\bibitem[{Fr{\"u}hwirth-Schnatter(1994)}]{fruhwirth1994data}
Fr{\"u}hwirth-Schnatter S (1994) Data augmentation and dynamic linear models.
\newblock \emph{Journal of Time Series Analysis} 15(2), 183--202

\bibitem[{Fr{\"u}hwirth-Schnatter and Wagner(2010)}]{fruhwirth2010stochastic}
Fr{\"u}hwirth-Schnatter S and Wagner H (2010) Stochastic model specification
  search for Gaussian and partial non-Gaussian state space models.
\newblock \emph{Journal of Econometrics} 154(1), 85--100

\bibitem[{Gallant et~al.(1997)Gallant, Hsieh and
  Tauchen}]{gallant1997estimation}
Gallant AR, Hsieh D and Tauchen G (1997) Estimation of stochastic volatility
  models with diagnostics.
\newblock \emph{Journal of Econometrics} 81(1), 159--192

\bibitem[{George et~al.(2008)George, Sun and Ni}]{george2008bayesian}
George EI, Sun D and Ni S (2008) Bayesian stochastic search for {VAR} model
  restrictions.
\newblock \emph{Journal of Econometrics} 142(1), 553--580

\bibitem[{Geweke(1994)}]{geweke1994bayesian}
Geweke J (1994) [Bayesian Analysis of Stochastic Volatility Models]: Comment.
\newblock \emph{Journal of Business \& Economic Statistics} 12(4), 397--399

\bibitem[{Geweke and Amisano(2010)}]{geweke2010comparing}
Geweke J and Amisano G (2010) Comparing and evaluating Bayesian predictive
  distributions of asset returns.
\newblock \emph{International Journal of Forecasting} 26(2), 216--230

\bibitem[{Geweke and Tanizaki(2001)}]{geweke2001bayesian}
Geweke J and Tanizaki H (2001) Bayesian estimation of state-space models using
  the Metropolis-Hastings algorithm within Gibbs sampling.
\newblock \emph{Computational Statistics \& Data Analysis} 37(2), 151--170

\bibitem[{Giannone et~al.(2015)Giannone, Lenza and
  Primiceri}]{giannone2015prior}
Giannone D, Lenza M and Primiceri GE (2015) Prior selection for vector
  autoregressions.
\newblock \emph{Review of Economics and Statistics} 97(2), 436--451

\bibitem[{Gordon and Smith(1990)}]{gordon1990modeling}
Gordon K and Smith A (1990) Modeling and monitoring biomedical time series.
\newblock \emph{Journal of the American Statistical Association} 85(410),
  328--337

\bibitem[{Griffin et~al.(2010)Griffin, Brown et~al.}]{griffin2010inference}
Griffin JE, Brown PJ et~al. (2010) Inference with normal-gamma prior
  distributions in regression problems.
\newblock \emph{Bayesian Analysis} 5(1), 171--188

\bibitem[{Harrison and Stevens(1976)}]{harrison1976bayesian}
Harrison PJ and Stevens CF (1976) Bayesian forecasting.
\newblock \emph{Journal of the Royal Statistical Society. Series B
  (Methodological)} , 205--247

\bibitem[{Huber and Feldkircher(2017)}]{huber2017adaptive}
Huber F and Feldkircher M (2017) Adaptive shrinkage in Bayesian vector
  autoregressive models.
\newblock \emph{Journal of Business \& Economic Statistics} , 1--13

\bibitem[{Huber et~al.(2017)Huber, Kastner and Feldkircher}]{huber2017new}
Huber F, Kastner G and Feldkircher M (2017) A New Approach Toward Detecting
  Structural Breaks in Vector Autoregressive Models.
\newblock \emph{arXiv preprint arXiv:607.04532v3}

\bibitem[{Jacquier et~al.(2004)Jacquier, Polson and
  Rossi}]{jacquier2004bayesian}
Jacquier E, Polson NG and Rossi PE (2004) Bayesian analysis of stochastic
  volatility models with fat-tails and correlated errors.
\newblock \emph{Journal of Econometrics} 122(1), 185--212

\bibitem[{Kastner(2015{\natexlab{a}})}]{kastner2015dealing}
Kastner G (2015{\natexlab{a}}) Dealing with stochastic volatility in time
  series using the R package stochvol.
\newblock \emph{Journal of Statistical Software. URL http://cran. r-project.
  org/web/packages/stochvol/vignettes/article. pdf}

\bibitem[{Kastner(2015{\natexlab{b}})}]{kastnerjss}
Kastner G (2015{\natexlab{b}}) Dealing with stochastic volatility in time
  series using the {R} package stochvol.
\newblock \emph{Journal of Statistical Software} forthcoming

\bibitem[{Kastner(2015{\natexlab{c}})}]{kastner2015heavy}
Kastner G (2015{\natexlab{c}}) Heavy-tailed innovations in the R package
  stochvol.
\newblock \emph{ePubWU Institutional Repository}

\bibitem[{Kastner and Fr{\"u}hwirth-Schnatter(2014)}]{kastner2014ancillarity}
Kastner G and Fr{\"u}hwirth-Schnatter S (2014) Ancillarity-sufficiency
  interweaving strategy (ASIS) for boosting MCMC estimation of stochastic
  volatility models.
\newblock \emph{Computational Statistics \& Data Analysis} 76, 408--423

\bibitem[{Markowitz(1952)}]{markowitz1952}
Markowitz H (1952) Portfolio Selection.
\newblock \emph{The Journal of Finance} 7(1), 77--91

\bibitem[{Primiceri(2005)}]{primiceri2005time}
Primiceri GE (2005) Time varying structural vector autoregressions and monetary
  policy.
\newblock \emph{The Review of Economic Studies} 72(3), 821--852

\bibitem[{Urquhart(2017)}]{urquhart2017price}
Urquhart A (2017) Price clustering in Bitcoin.
\newblock \emph{Economics Letters} 159, 145--148

\bibitem[{West(1987)}]{west1987scale}
West M (1987) On scale mixtures of normal distributions.
\newblock \emph{Biometrika} 74(3), 646--648

\end{thebibliography}
\addcontentsline{toc}{section}{References}

\begin{appendices}

\end{appendices}

\end{document}